# Dendronized Mesoporous Silica Nanoparticles Provide an Internal Endosomal Escape Mechanism for Successful Cytosolic Drug Release

*Veronika Weiss,[a] Christian Argyo,[a] Adriano A. Torrano,[a] Claudia Strobel,[b] Stephan A. Mackowiak,[a] Tim Gatzenmeier,[a] Ingrid Hilger,[b] Christoph Bräuchle,[a]\* and Thomas Bein [a]\**

[a] Department of Physical Chemistry and Center of NanoScience (CeNS), University of Munich (LMU), Gerhard-Ertl-Building, Butenandtstraße 5-13, 81377 München (Germany)

[b] Department of Experimental Radiology, Institute of Diagnostic and Interventional Radiology I, Jena University Hospital, Friedrich-Schiller-University Jena, Erlanger Allee 101, 07747 Jena (Germany)

† V. Weiss and C. Argyo contributed equally to this work

\* E-mail: bein@lmu.de, christoph.braeuchle@cup.uni-muenchen.de

**Abstract**

Mesoporous silica nanoparticles (MSNs) attract increasing interest in the field of gene and drug delivery due to their versatile features as a multifunctional drug delivery platform. Here, we describe poly(amidoamine) (PAMAM) dendron-functionalized MSNs that fulfill key prerequisites for a controllable intracellular drug release. In addition to high loading capacity, they offer 1) low cytotoxicity, showing no impact on the metabolism of endothelial cells, 2) specific cancer cell targeting due to receptor-mediated cell uptake, 3) a redox-driven cleavage of disulfide bridges allowing for stimuli-responsive cargo release, and most importantly, 4) a specific internal trigger based on the high buffering capacity of PAMAM dendrons to provide endosomal escape.

## 1. Introduction

In recent years, mesoporous silica nanoparticles (MSNs) have been intensively studied as drug delivery vehicles, due to their excellent material features such as good biocompatibility, large cargo capacity, and versatile organic surface functionalization [1].

In general, an ideal drug delivery platform has to meet several requirements to achieve specific drug delivery [2-4], including biocompatibility [5-7], specific targeting [8, 9], and stimuli-responsive drug release behavior [10-12]. Nano-sized drug delivery systems such as multifunctional MSNs encounter many challenges on their way towards reaching their desired target and efficiently releasing their cargo. In particular, the endosomal entrapment is a major obstacle for drug delivery, faced by MSNs that are internalized by cells via endocytosis [13]. Especially for membrane impermeable or immobilized cargo molecules, the nanocarriers must enter the cytosol to achieve efficient delivery to the targeted cell compartments. Several strategies have already been described to address the demanding task of endosomal escape, including pore formation, membrane fusion, photoactivated membrane rupture, and the proton sponge effect [14-17].

The latter is supposed to be a promising automatic strategy for endosomal release of the nanocarriers. The mechanism of the proton sponge effect follows an intrinsic osmotic swelling during endosomal acidification caused by the buffering capacity of modified nanocarriers such as cationic polymers [18-20]. Ultimately, this results in rupture of the endosomal membrane. Furthermore, a destabilization of the membrane caused by such positively charged vehicles has been proposed [21].

Poly(amidoamine) (PAMAM) dendrons or dendrimers provide high buffering capacity and have been found to be suitable for gene delivery exhibiting extraordinary stability in forming complexes with DNA [22-25]. Resulting transfection efficiency was explicitly attributed to an activated proton sponge mechanism. Dendron-coated mesoporous particles have also been used for intracellular delivery of plasmid-DNA [26, 27].

Here we establish newly designed multifunctional core-shell MSNs coated with PAMAM dendron structures on the outer surface. These systems provide a successful mechanism for endosomal escape and subsequent cytosolic drug release from the silica nanocarriers. PAMAM dendron-coated MSNs feature a high buffering capacity acting as a potential trigger for a pH-responsive endosomal escape mechanism conceivably via the proton sponge effect (Scheme 1). To the best of our knowledge, we created for the first time a multifunctional drug delivery platform based on the combination of polycationic PAMAM dendrons as endosomal escape agents with a redox-responsive drug release mechanism (disulfide bridges inside mesopores) and the possibility to attach cancer cell targeting ligands to the MSN nanocarriers.

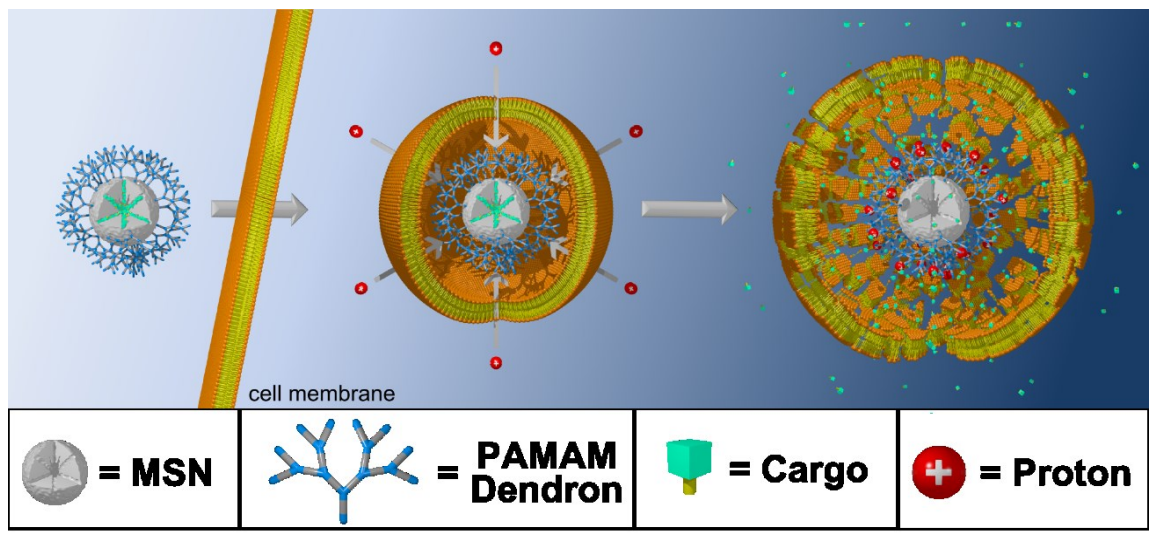

**Scheme 1.** Schematic illustration of the proposed intrinsic endosomal escape mechanism. PAMAM dendron-coated mesoporous silica nanoparticles are internalized into a cancer cell via endocytosis. Endosomal acidification leads to intrinsic osmotic swelling caused by the high buffering capacity of the dendron-coated MSNs. Subsequently, endosomal membrane rupture occurs which provides access to the cytosol. In this reductive environment the immobilized (disulfide bridges) cargo molecules can be released.

## 2. Experimental

**Materials and methods**

**Materials.** Propargylamine (Aldrich, 98 %), *N*,*N*-diisopropylethylamine (DIPEA, Sigma-Aldrich, ≥ 99 %), copper(I) iodide (Aldrich, 99.999 %), (3-aminopropyl) triethoxysilane (APTES, Aldrich, 99%), tetraethyl orthosilicate (TEOS, Fluka, > 98 %), triethanolamine (TEA, Aldrich, 98 %), cetyltrimethylammonium chloride (CTAC, Fluka, 25 % in $H_2O$), (3-mercaptopropyl) trimethoxysilane (MPTMS, Gelest, 95 %), methanethiosulfonate 5(6)-carboxy-X-rhodamine (MTS-ROX, Biotium), colchicine methanethiosulfonate (MTS-Col, Santa Cruz Biotechnology), 4,6-diamidino-2-phenylindole dihydrochloride (DAPI, Sigma Aldrich), folic acid (FA, Sigma), Atto 633 maleimide (ATTO-TEC), 1-ethyl-3-(3-dimethylaminopropyl) carbodiimide hydrochloride (EDC, Fluka, 97 %), *N*-hydroxysulfosuccinimide (sulfo-NHS, Sigma Aldrich, >98.5 %), poly (ethylene glycol) bisamine ($PEG_{2000}$-bis$NH_2$, $M_W$ 2000, Sigma Aldrich), *N*-succinimidyl oxycarbonylethyl methanethiosulfonate (NHS-3-MTS, Santa Cruz Biotechnology), and oxalic acid were used as received. Methyl acrylate (Aldrich, 99 %) and 1,2-ethylendiamine (Aldrich, 99.5 %) were freshly distilled prior to use. Ethanol (EtOH, absolute), *N*,*N*-dimethyl formamide (DMF, Sigma Aldrich, anhydrous) and methanol (MeOH, anhydrous, Sigma) were used as solvents without further purification. Bidistilled water was obtained from a Millipore system (Milli-Q Academic A10). (3-azidopropyl)trimethoxy silane (AzTMS) was freshly prepared as previously reported [28].

**Dendronized MSNs (MSN-D3).** Colloidal mesoporous silica nanoparticles (MSN-D3) were prepared according to a synthesis procedure published by Cauda *et al.* [29]. The amount of functionalized silane was calculated to be 1% of total silica. A mixture of TEA (14.3 g, 95.6 mmol), TEOS (1.56 g, 7.48 mmol) and MPTMS (92.3 mg, 87.3 mL, 0.47 mmol) was heated for 20 min under static conditions at 90 °C in a polypropylene reactor. Afterwards, a solution of cetyltrimethylammonium chloride (CTAC, 2.41 mL, 1.83 mmol, 25 wt% in $H_2O$) in $H_2O$ (21.7 g, 1.21 mmol) was preheated to 60 °C and added quickly to the TEOS solution. The reaction mixture was stirred vigorously (700 rpm) for 20 min while cooling down to room temperature. Subsequently, TEOS (138.2 mg, 0.922 mmol) was added in four equal increments every three minutes. The reaction was stirred for further 30 min. After this time, a mixture of TEOS (19.2 mg, 92.2 µmol) and a dendron-functionalized trialkoxysilane (S3, cf. SI) (92.2 µmol) was added. Furthermore, the mixture of TEOS and PAMAM silane was dissolved in a solution of 2 mL methanol and 1 mL water briefly before the addition. The reaction was stirred at room temperature overnight. The suspension was diluted 1:1 with absolute ethanol, the colloidal MSNs were separated by centrifugation (19,000 rpm, 43,146 rcf, 20 min) and redispersed in absolute ethanol. The template extraction was performed by heating the samples under reflux at 90 °C (oil bath) for 45 min in a solution of ammonium nitrate (2 wt% in ethanol) followed by 45 min under reflux at 90 °C in a solution of 10 mL conc. HCl (37 %) in 90 mL ethanol. The extracted MSNs were collected by centrifugation after each extraction step and finally washed with 100 mL absolute ethanol. The resulting MSNs were stored in an ethanol/water solution (2:1).

**Cargo loading.** MSNs (1 mg) were incubated for 2 h at room temperature in the dark with MTS-ROX (5 µL, 5 mg/mL in DMF), MTS-Col (50 µL, 5 mg/mL in DMF), or MTS-DAPI for a one-step covalent attachment of the cargo molecules to the internal surface of the mesopores via disulfide bridges. MTS-DAPI was prepared in-situ by mixing DAPI (10 µL, 5 mg/mL in DMF) and NHS-3-MTS (100 µg) in an aqueous solution for 1 h at room temperature in the dark. This reaction mixture was subsequently added to MSN-D3. The particles were washed five times (centrifugation, 4 min, 14,000 rpm, 16,837 rcf) with water and were finally redispersed in 1 mL $H_2O$.

**MSN labeling.** MSNs (1 mg) were mixed with Atto 633 maleimide (1 µL, 2 mg/mL in DMF). After 2 h of stirring at room temperature in the dark the labeled particles were washed three times with water by subsequent centrifugation (4 min, 14,000 rpm, 16,837 rcf) and redispersed in $H_2O$.

**Attachment of the targeting ligand folate.** Oxalic acid, EDC, and sulfoNHS (each 5 mg) were successively added to an aqueous dispersion of MSN-D3 (0.5 mg) and the resulting reaction mixture was stirred at room temperature in the dark for another 2 h. Subsequently, the particles were washed three times (centrifugation, 4 min, 14,000 rpm, 16,837 rcf) and redispersed in 300 µL $H_2O$. The PEG linker was attached via EDC amidation. EDC, sulfoNHS and $PEG_{2000}$-$bisNH_2$ (each 5 mg) were consecutively added to the particle dispersion and stirred for another 2 h at room temperature in the dark. The particles were washed three times by centrifugation (4 min, 14.000 rpm) and redispersed in 100 µL water. Then 400 µL of a folate stock solution (0.75 µM in $H_2O$), EDC (5 mg), and sulfoNHS (5 mg) were added. The reaction mixture was stirred overnight at room temperature in the dark. After washing the particles three times with water (centrifugation, 4 min, 14,000 rpm, 16,837 rcf) the MSN-D3-FA sample was finally redispersed in 1 mL $H_2O$.

**Characterization.** NMR spectra were recorded on a Jeol Eclipse 270 ($^1$H: 270 MHz, $^{13}$C: 67.9 MHz), a Jeol Eclipse 400 ($^1$H: 400 MHz, $^{13}$C: 101 MHz) NMR or a Jeol 500 Eclipse spectrometer ($^1$H: 500.16 MHz, $^{13}$C: 125.77 MHz). $^{13}$C solid-state NMR measurements were performed on a Bruker DSX Avance500 FT spectrometer in a 4 mm $ZrO_2$ rotor. IR spectra were recorded on a Bruker Equinox 55. The dried powder of the silica nanoparticles (1.5 mg) was mixed with KBr (200 mg) to produce pressed, transparent pellets to be measured in absorbance mode. For background measurements a neat KBr pellet (200 mg) was used. IR spectra of the organic compounds were recorded in the attenuated total reflectance (ATR) mode on the same devise or on a Perkin-Elmer FT-IR Spektrum BXII spectrometer with Schmith Dura SamplIR II ATR-Unit. Mass spectra were measured on a Thermo Finnigan LTQ FT with IonMax (ion source with ESI head). Elementary analysis was performed on a Vario EL or Vario micro cube CHN analyzer detecting carbon, nitrogen, hydrogen and sulfur. Acid-base titrations were performed on a Metrohm 905 Titrando potentiometric titrator combined with the software tiamo. The titration method met the following parameters: monotonic equivalence point titration (MET); stirring speed 5; measurement value drift 20.0 mV/min; min. delay 0 s; max. delay 300 s; volume increment 0.025 mL; dosing speed maximal; stop value pH 10.2. The samples were prepared as follows: A volume containing 13.7 mg particles from the suspensions was added to 30.0 mL $H_2O$. The starting pH was set to 3.000 using HCl (0.1 M) and NaOH (0.01 M) from the dosing unit. The samples were titrated against NaOH (0.01 M). TGA of the bulk extracted samples (about 10 mg of dried powder) were performed on a Netzsch STA 440 Jupiter thermobalance (heating rate of 10 K/min in a stream of synthetic air of about 25 mL/min). Nitrogen sorption measurements were performed on a Quantachrome Instruments NOVA 4000e at -196 °C. Sample out-gassing was performed at 120 °C for 12 h at a pressure of 10 mTorr. Pore size and pore volume were calculated using a NLDFT equilibrium model of $N_2$ on silica, based on the desorption branch of the isotherm. The BET model was used in the range of 0.05 – 0.20 $p/p_0$ to estimate the specific surface area. DLS and zeta potential measurements were performed on a Malvern Zetasizer-Nano instrument equipped with a 4 mW He-Ne laser (633 nm) and avalanche photodiode detector. DLS measurements were directly recorded on aqueous colloidal suspension at a constant concentration for all sample solutions of 1 mg/mL. For determination of the zeta potential profiles, one to three drop of the ethanolic suspension (ca. 3 wt%) was mixed with 1 mL of commercial Hydrion Buffer solution of the appropriate pH prior to measurement. Hückel's approximation was used for interpretation. For TEM (using a Titan 80–300 kV microscope operating at 300 kV), samples were prepared by dispersing MSNs (1 mg) in 4 mL absolute ethanol, by means of an ultrasonic bath, and drying a drop of the resulting diluted suspension on a carbon-coated copper grid. Fluorescence spectra were recorded on a PTI spectrofluorometer equipped with a xenon short arc lamp (UXL-75XE USHIO) and a photomultiplier detection system (model 810/814). The measurements were performed in aqueous solution at 37 °C to simulate human body temperature. For time-based release experiments of MTS-ROX a custom made container consisting of a Teflon tube, a dialysis membrane (ROTH Visking type 8/32, MWCO 14,000 g/mol) and a fluorescence cuvette was used. The excitation wavelength was set to $\lambda$ = 575 nm for MTS-ROX-loaded MSNs. Emission scans (585 – 650 nm) were performed every 2 min. All slits were adjusted to 1.0 mm, bandwidth 8 nm).

**Cytotoxicity studies.** The cytotoxic impact of the PAMAM dendron-coated MSNs was determined by using immortalized human microvascular endothelial cells (HMEC-1; Centers for Disease Control and Prevention, USA). HMEC-1 were cultivated in Gibco® MCDB 131 medium (Life Technologies GmbH, Germany) supplemented with 10 % (v/v) fetal bovine serum (FBS, Life Technologies GmbH, Germany), 1 % (v/v) GlutaMAX$^{TM}$ I 100X (Life Technologies GmbH, Germany), 1 µg/ml hydrocortisone (Sigma-Aldrich Chemie GmbH, Germany) and 10 ng/ml epidermal growth factor (Life Technologies GmbH, Germany) at 37 °C in a 5 % $CO_2$ humidified environment by changing the growth medium every 2-3 days. Subcultivation was performed after reaching cellular confluency of 70 – 85 % using GIBCO® trypsin (Life Technologies GmbH, Germany) for cell detachment. PCR was routinely performed to test that the cells were free of mycoplasma infection. For cytotoxicity evaluation, 12.000 cells/cm$^2$ were seeded in a 96 well cell culture plate. 24 h after seeding they were exposed to PAMAM dendron-coated MSNs of different concentrations (10 fg/ml to 100 µg/ml) for 3 to 72 h. After the defined incubation times, the cells were washed with Hank`s BSS and freshly cell culture medium and 20 µl Cell titer 96 Aqueous One Solution Reagent (Promega GmbH, Germany) per well were added. The absorbance of the supernatants was measured at 492 nm via a microplate reader (Sunrise™, Tecan Group Ltd., Switzerland) and the relative cellular dehydrogenase activity of endothelial cells was calculated by normalizing the values to untreated control cells. For data interpretation the cytotoxic threshold given by DIN EN ISO 10993-5:2009-10 was used.

For evaluation of nanoparticle uptake kinetics, 12.000 cells/cm$^2$ were seeded in 8-well LabTek-II slides (Nunc). Cells were exposed to 100 µg/ml of PAMAM dendron-coated MSNs 24 h after seeding for 3, 24 and 48 h. After that, cells were washed twice with phosphate buffered and plasma membrane was stained with a 10 µg/ml wheat germ agglutinin Alexa Fluor 488 conjugate (WGA488, Invitrogen) solution in cell medium during 1 minute. Staining solution was removed and cells were washed twice with cell medium. Fresh cell medium was added and cells were imaged immediately by spinning disc confocal microscopy (as described below) with a Z-spacing of 0.25 µm (63x objective). In order to reach an absolute quantification of nanoparticle, the intensity of each object (nanoparticle or agglomerate) is compared to the intensity of a single nanoparticle previously measured in a calibration procedure. Calibration experiments were carried out with PAMAM dendron-coated MSNs deposited on a cover slip and covered with cell medium. The images were evaluated with the subroutine 'Calibration' of Particle_in_Cell-3D and the mean intensity value showed a Gaussian distribution with the mean at 64632 pixel intensities per nanoparticle.

**Cell culture.** KB cells were grown in folic acid deficient Roswell Park Memorial Institute 1640 medium (RPMI 1640, Invitrogen) supplemented with 10 % fetal bovine serum (FBS) at 37 °C in a 5 % $CO_2$ humidified atmosphere. The cells were seeded on ibidiTreat µ-Slide (IBIDI). HeLa cells were grown in Dulbecco's modified Eagle's medium (DMEM):F12 (1:1) (Invitrogen) with Glutamax I medium supplemented with 10 % FBS at 37 °C in a 5 % $CO_2$ humidified atmosphere. The cells were seeded on collagen A-coated LabTek chambered cover glass (Nunc). For live cell imaging the cells were seeded 24 or 48 h before measuring, at a cell density of $2x10^4$ or $1x10^4$ cells/cm$^2$.

*In vitro* **cargo release.** Cells were incubated 5 – 48 h prior to the measurements at 37 °C under a 5% $CO_2$ humidified atmosphere. Shortly before imaging, the medium was replaced by $CO_2$-independent medium (Invitrogen). During the measurements all cells were kept on a heated microscope stage at 37 °C. The subsequent imaging was performed as described in the Spinning disk confocal microscopy section.

**Nuclei staining kinetics with DAPI.** HeLa cells were measured 5, 11, 24, 33, 49, and 61 h after incubation with the samples MSN-D3-MTS-DAPI, MSN-$NH_2$-MTS-DAPI and the supernatant of MSN-D3-MTS-DAPI (after particle separation). Each time point was measured with an independently incubated set of cells. In order to evaluate the fluorescence of the nucleoli, the z-stack position was set exclusively in the region of the nuclei. The integrated intensity of a distinct region of interest (ROI) around the nuclei was determined (44 – 104 nuclei for each data point) and divided by the area of the ROI. Consequently, the average integrated density per nuclei area (± standard deviation, represented by error bars) could be plotted dependent on incubation time. The relatively large error bars are supposed to result from a different particle uptake behavior for each cell. The amount of uptaken particles to the cells was not taken into account.

**Cell targeting.** To evaluate the functionality of the folic acid ligand, KB cells were incubated with nanoparticles for 5 h at 37 °C under a 5 % $CO_2$ humidified atmosphere. The cell membrane was stained shortly before the measurement by adding 4 µL of 1 mg/mL wheat germ agglutinin Alexa Fluor 488 conjugate (WGA488, Invitrogen) to 400 µL of cell medium. After 1 min, the cell medium was removed, the cells were washed twice with cell medium, and imaged immediately. In control experiments, the FA receptors on the KB surface were blocked by pre-incubation of the cells with 3 mM folic acid (Sigma) for 2 h at 37 °C under a 5% $CO_2$ humidified atmosphere, before particles were added.

Spinning disc confocal microscopy. Confocal microscopy for live-cell imaging was performed on a setup based on the Zeiss Cell Observer SD utilizing a Yokogawa spinning disk unit CSU-X1. The system was equipped either with a 1.40 NA 100x or a 1.40 NA 63x Plan apochromat oil immersion objective from Zeiss. For all experiments the exposure time was 0.1 s and z-stacks were recorded. WGA488 was imaged with approximately 0.4 W/mm$^2$ and GFP with 1.3 W/mm$^2$ of 488 nm excitation light. Atto 633 was excited with 0.12 W/mm$^2$ of 639 nm and DAPI with 0.2 W/mm$^2$ of 405 nm. In the excitation path a quad-edge dichroic beamsplitter (FF410/504/582/669-Di01-25x36, Semrock) was used. For two color detection of WGA488, GFP, or DAPI and Atto 633, a dichroic mirror (560 nm, Semrock) and band-pass filters 525/50 and 690/60 (both Semrock) were used in the detection path. Separate images for each fluorescence channel were acquired using two separate electron multiplier charge coupled devices (EMCCD) cameras (PhotometricsEvolveTM).

## 3. Results and discussion

### 3.1 Preparation of dendronized MSNs

MSNs coated with third generation PAMAM dendrons (MSN-D3) were synthesized via a delayed co-condensation approach to create core-shell functionalized MSNs [29]. Specifically, bifunctional MSNs consist of a thiol-functionalized particle core and additionally an external PAMAM dendron shell (MSN-D3).

Particle size and structural parameters derived by dynamic light scattering and nitrogen sorption measurements, respectively, are summarized in **Table 1**. Nitrogen sorption measurements showed a type IV isotherm indicating a mesoporous structure of the nanoparticles (cf. SI). Attachment of the dendrons decreased the porosity parameters which is attributed to several factors such as pore mouth narrowing (see discussion in SI). Dynamic light scattering measurements confirmed small particle sizes with a narrow particle size distribution (Table 1 and Figure S2a). Transmission electron micrographs (Figure S2c) show spherically shaped nanoparticles with sizes of about 70 nm in diameter for MSN-D3. Furthermore, a worm-like porous structure consisting of radially grown mesoporous channels was observed for all samples (cf. SI). Zeta potential measurements showed drastic changes in the surface charge of MSN-D3 compared to MSNs without PAMAM dendron functionality (MSN-D0). As depicted in **Figure 1a**, highly positive surface charges were observed at acidic pH values for MSN-D3 (+60 mV at pH 2). In contrast, the reference sample MSN-D0 exhibited an isoelectric point close to pH 2 which results in a negatively charged particle surface over the full pH range. Additional titration measurements of MSN-D3 against an aqueous solution of NaOH (0.01 M) gave evidence for a high proton acceptance of the polymer shell resulting in a high buffering capacity. As depicted in Figure 1b, MSN-D3 featured a significant increase in required volume of NaOH solution to be neutralized (+ 1.4 mL). MSN-D3 provided great potential to act like a proton sponge showing optimal buffering behavior in the pH range 5.5 to 6.5 which perfectly fits the endosomal acidification range.

Further characterization of the attached functional groups was performed by $^{13}$C solid state NMR analysis (Figure 1c, for more details see the SI). MSN-D3 featured characteristic peaks for the amide groups of the PAMAM dendrons at 173 ppm (C=O). Furthermore, weak signals at 144 and 125 ppm were derived from the triazole click connection. Various strong signals in the range between 60 to 10 ppm correspond to different types of methylene groups which belong to the PAMAM moieties (52 and 40 ppm (N-CH$_2$-R), 21 and 10 ppm (R-CH$_2$-R)). From all of these results (cf. SI for further characterization), we conclude a successful synthesis of core-shell functionalized MSNs with third generation PAMAM dendrons via the delayed co-condensation approach. The PAMAM moieties are exclusively located at the external particle surface resulting in an organic polymer coating of MSNs featuring a high buffering capacity.

**Table 1.** Structural parameters of functionalized MSNs.

| Sample | Particle size[a] [nm] | BET surface area [m²/g] | Pore volume[b] [cm³/g] | DFT pore size[c] [nm] |
|---|---|---|---|---|
| MSN-D0 | 122 | 1190 | 0.74 | 2.9 – 4.4 |
| MSN-D3 | 122 | 497 | 0.24 | 2.2 – 3.8 |

[a]Particle size is given by the hydrodynamic diameter and refers to the peak value of the size distribution derived from DLS measurements; [b]Pore volume is calculated up to a pore size of 8 nm to remove the contribution of the interparticle porosity; [c]DFT pore size refers to FWHM of the corresponding pore size distribution.

### 3.2 Redox-driven cargo release

To prove a stimuli-responsive cargo release behavior of MSN-D3 we performed time-based release experiments of a fluorescent and thiol-reactive model drug (MTS-ROX). MSN-D3 with immobilized MTS-ROX in the mesopores showed no cargo release in aqueous solution under non-reductive conditions within the first hour (Figure 1d). The cargo molecules were covalently attached via disulfide bridges preventing premature leakage of the dye. Only upon addition of a reducing agent (simulation of the cytosol) an increase in fluorescence intensity was shown. This indicated a redox-responsive cleavage of the disulfide bridges and subsequently a specific stimuli-responsive release of the fluorescent model drug occurred. In conclusion, the redox-responsive behavior provides great potential for a specific cargo release once the mesoporous silica nanocarriers have escaped from the endosomes and entered the cytosol of the targeted cell.

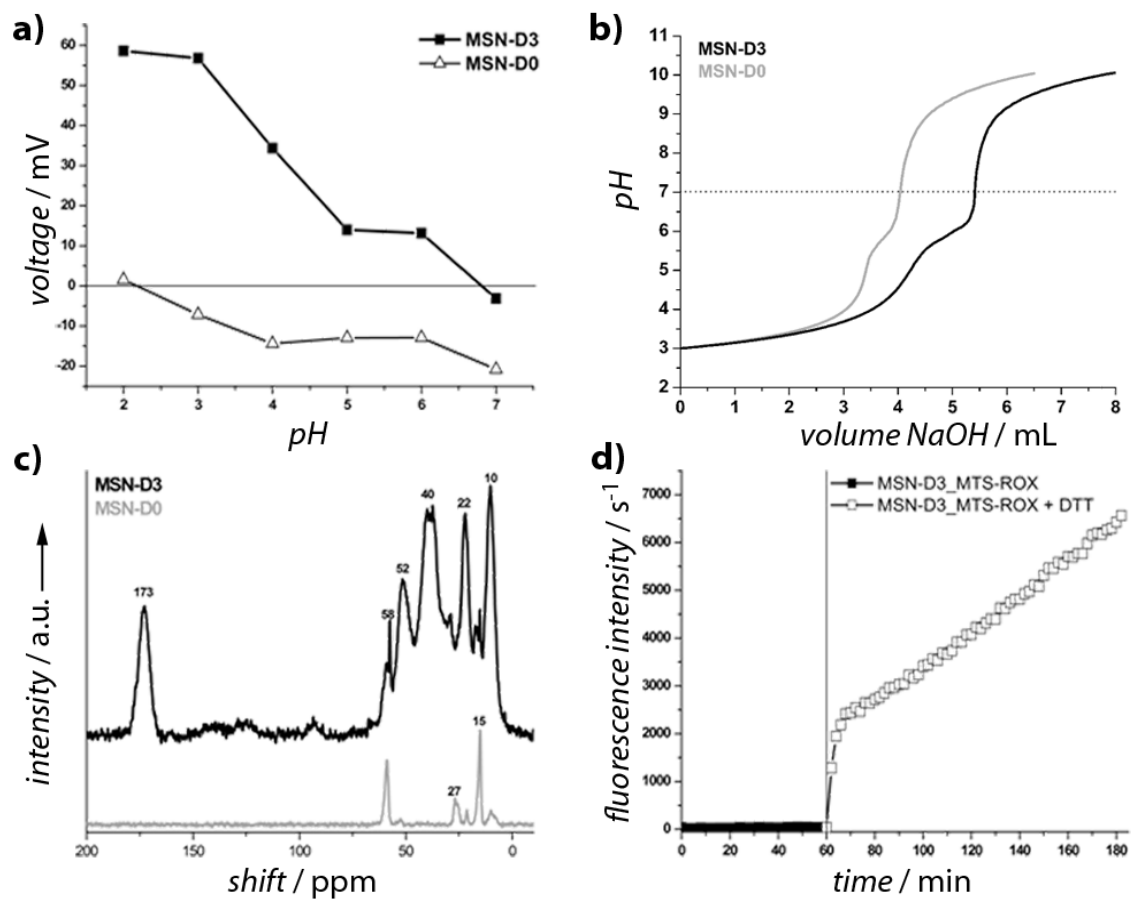

**Figure 1.** Characterization of PAMAM dendron-coated MSNs. a) Zeta potential measurements, b) titration data and c) solid-state NMR measurements (for clarity the graphs have been shifted along the y-axis) of MSN-D3 and MSN-D0. d) Redox-responsive release kinetics of MTS-ROX before (filled squares) and after (empty squares) addition of dithiothreitol (DTT) to simulate the reductive milieu of the cytosol. Before medium exchange, MSN-D3 shows no premature release of the fluorescent cargo molecules which are attached to the mesopores via disulfide bridges. Only in reductive milieu a significant increase in fluorescent intensity can be observed, demonstrating a redox-responsive release behavior of the MTS-ROX due to cleavage of the disulfide bridges.

### 3.3 Cytotoxicity studies

In order to elucidate the toxicity of PAMAM dendron-coated MSNs we determined the uptake kinetics and the relative cellular dehydrogenase activity of immortalized human microvascular endothelial cells (HMEC-1), modeling an important cell barrier after intravenous application (**Figure 2**). First, to gain detailed information about the dose of MSNs effectively internalized by cells at defined incubation times, we investigated the nanoparticle uptake kinetics (Figure 2d). HMEC-1 cells were exposed to a concentration of 100 µg/mL MSN-D3 from 3 to 48 h and subsequently imaged by live-cell fluorescence microscopy. Confocal stack images of single cells interacting with fluorescent-labeled nanoparticles were acquired and evaluated with Particle_in_Cell-3D, a specifically developed ImageJ image analysis method [30]. Image analysis revealed that around 2,500 nanoparticles were inside each cell after just 3 h. This number reached more than 13,000 nanoparticles per cell after 24 h. Interestingly, the intracellular particles are "diluted" by the natural cell division process [31], decreasing to approximately 10,000 MSNs per cell after 48 h. This observation is in accordance with the reported doubling time of HMEC-1 cells (29 h) [32]. The relative cellular dehydrogenase activity (rcDH) of HMEC-1 cells after exposure to MSN-D3 is presented in Figure 2e. Although the cells were shown to become packed with thousands of nanoparticles, MSN-D3 revealed no adverse effects over all investigated concentrations (10 fg/mL to 100 µg/mL) and exposure times (3 to 72 h). Therefore, PAMAM dendron-coated MSNs have a low cytotoxicity, comparable to other PAMAM-coated materials, as previously described [33, 34].

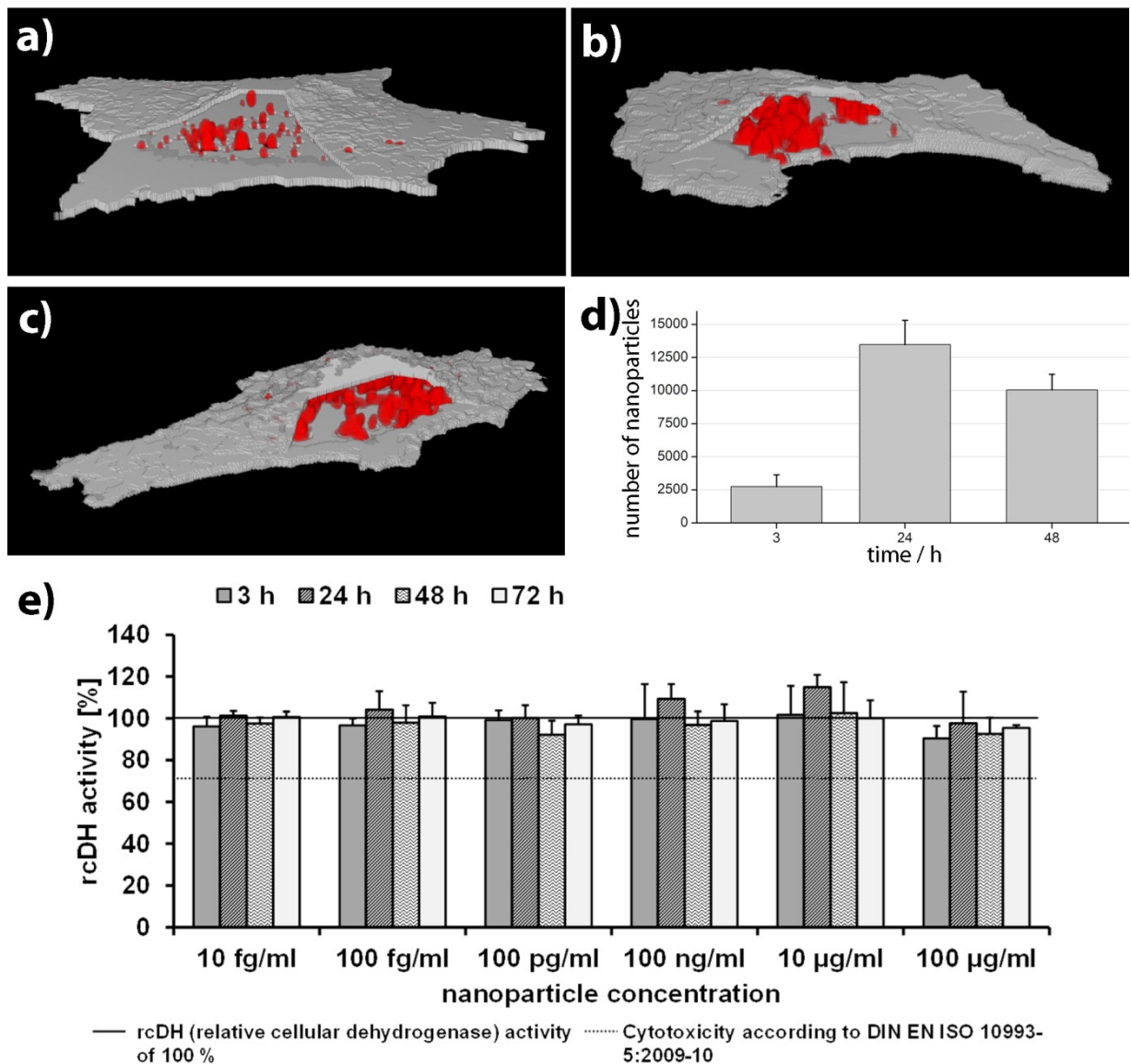

**Figure 2.** Cellular uptake kinetics and cytotoxicity studies of PAMAM dendron-coated MSNs. Panels a-c) show representative 3D images of single cells (gray) and intracellular MSN-D3 (red) after 3 h, 24 h and 48 h, respectively. d) The absolute number of nanoparticles internalized by single cells increases steeply within the first 24 h. Remarkably, the number slightly decreases after 48 h. The histogram depicts the mean ± standard error of two independent experiments (n=15). e) Relative cellular dehydrogenase activity (rcDH) of human microvascular endothelial cells (HMEC-1) after exposure to particles. MSN-D3 revealed no cytotoxic impact (n=3 independent experiments).

### 3.4 Specific receptor-mediated cell uptake

Another key feature for a capable drug delivery vehicle is specific targeting of the desired tissue, mainly cancer cells, which can be achieved by exploiting a receptor-mediated cellular uptake of the nanocarriers. Targeting ligands, such as folic acid (FA), can be attached to the periphery of the MSN-D3 via a bifunctional PEG linker (cf. SI). Application of the PEG$_{2000}$-linker is considered to reduce unspecific cellular uptake of the particles [35]. Atto 633-labeled MSN-D3 (red) equipped with PEG-FA were incubated for 5 h with KB cells to investigate a receptor-mediated endocytosis. Two cellular uptake experiments were performed simultaneously (**Figure 3**). On the one hand, KB cells were pretreated with free FA to gain full saturation and blocking of the cell receptors before particle incubation. Here, only unspecific cell uptake could be observed to a minor degree. A significant higher efficiency of particle internalization to the cells occurred for KB cells without pretreatment. Here, the fast cellular uptake is attributed to the receptor-mediated endocytosis of MSNs with FA.

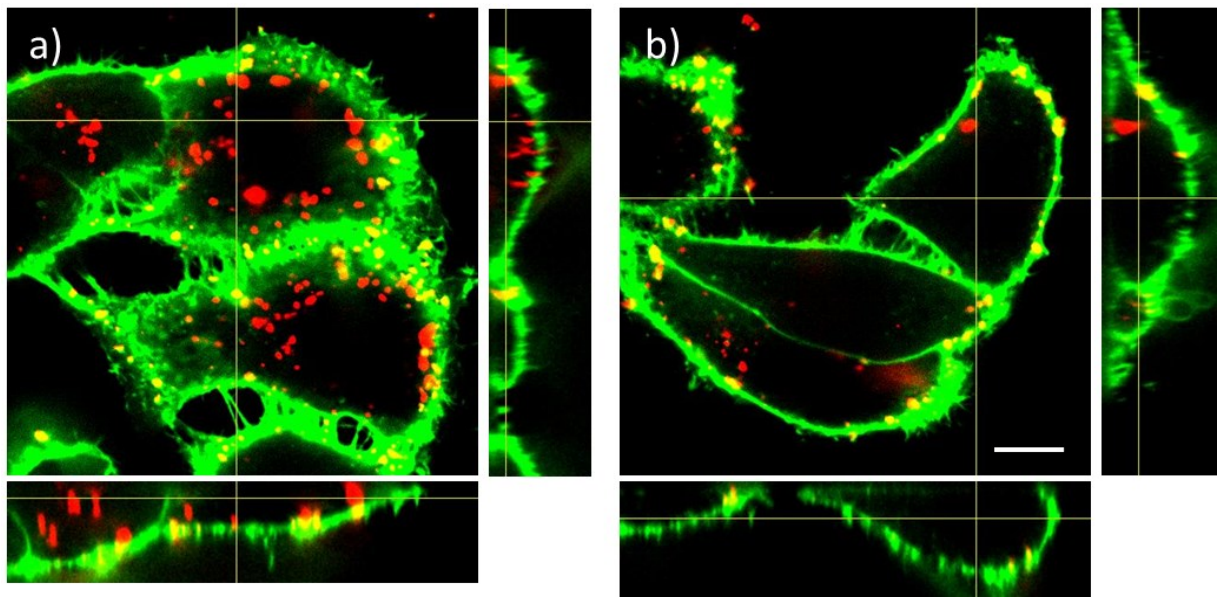

**Figure 3.** a) Receptor-mediated and b) unspecific endocytosis of MSN-D3 with targeting ligand folate (MSN-D3-PEG-FA, red) by KB cells (membrane staining, green). Indication of the particle location (inside or outside the cell) can be received by evaluation of the cellular z-stacks with spinning disk confocal fluorescence microscopy which can also be seen in the orthogonal views. a) A specific receptor-mediated cell uptake can be observed for MSN-D3-PEG-FA with KB cells (not FA-preincubated) after 5 h incubation at 37 °C. b) Incubation of MSN-D3-PEG-FA with FA-preincubated KB cells for 5 h at 37 °C showed only minor unspecific cellular uptake. The scale bar represents 10 µm.

**3.5 Endosomal escape**

We investigated the potential of MSN-D3 to achieve an intrinsic endosomal escape by performing *in vitro* release experiments of the nuclei staining DAPI on HeLa cancer cells [36, 37]. A thiol-reactive derivative of DAPI (MTS-DAPI) was covalently attached to the mercapto-functionalized walls of the mesopores. The established covalent disulfide bridges should be cleaved only when endosomal escape was achieved and the nanocarriers have reached the reductive milieu of the cytosol [38, 39]. HeLa cells were incubated with MSN-D3 loaded with MTS-DAPI for 5, 10 and 22 h. **Figure 4** shows an efficient cellular uptake behavior of the nanoparticles labeled with Atto 633 (red) already after 5 h. Moreover, only weak staining of the nuclei (DAPI, blue) was observed at this time point. In contrast, free DAPI molecules are able to efficiently stain nuclei already within a few minutes (5 min), as described by standard nucleus staining protocols [40]. Over the entire range a successive increase in fluorescence intensity of the DAPI-stained nuclei could be observed (Figure 4d). We attribute this time-dependent release behavior of DAPI to the relatively slow process of the proton sponge effect which was caused by the PAMAM dendron content of our nanocarriers. Consequently, endosomal escape for the cytosolic delivery of the cargo molecules was provided. Of note, the MSNs remained stationary after the endosomal escape as has been observed in previous reports [16, 41]. Sample MSN-D3 shows a significant increase in fluorescence intensity monitored for a total time period of 61 h (Figure 4d). This proves an effective DAPI delivery to the cells shown by fluorescent nuclei staining. Reference samples (MSN-D0 and supernatant of MSN-D3 after particle separation by centrifugation, MSN-D3_supernatant) showed no temporal increase of fluorescence intensity proving that only marginal amounts of free dye were present in the solution. Most important, this proves that the PAMAM dendron content of our drug delivery vehicles is essential to achieve successful cargo delivery to cancer cells.

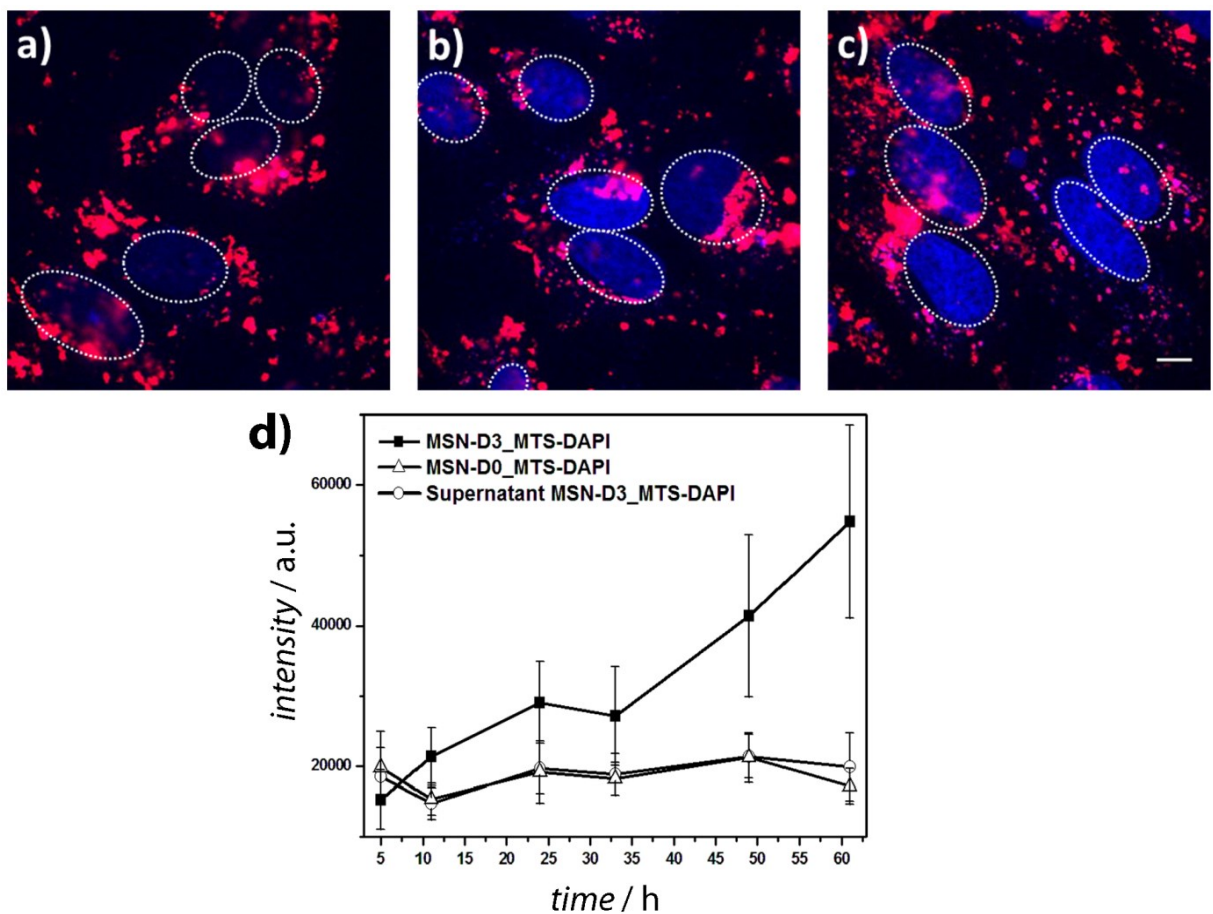

**Figure 4.** Fluorescence microscopy studies of HeLa cells incubated with MSN-D3 loaded with immobilized DAPI (MTS-DAPI, blue) inside the mesopores and labeled with Atto 633 (red) after a) 5 h, b) 10 h and c) 22 h. The nuclei are indicated with dashed circles. The scale bar represents 10 µm. d) Nuclei staining kinetics of DAPI delivery to HeLa cells from MSN-D3_MTS-DAPI (squares), MSN-D0_MTS-DAPI (triangles) and the supernatant (circles) of MSN-D3_MTS-DAPI solution (after particle separation). The fluorescence intensity of distinct regions of interest (stained nuclei) was evaluated after different time points of sample incubation. Data represents average fluorescence intensity ± standard deviation. A time-dependent increase of fluorescence intensity can be observed only for MSN-D3_MTS-DAPI suggesting a gradual DAPI release from the nanocarriers. In contrast, the fluorescence intensity remains constant at a marginal level for the reference samples.

The anticancer drug colchicine (Col) is known to cause inhibition of the microtubuli polymerization due to irreversible binding to tubulins ultimately leading to cell death [42, 43]. Here, a thiol-reactive derivative of Col (MTS-Col) was immobilized at the inner mesoporous surface of MSN-D3 and subsequently, these loaded drug delivery vehicles were incubated with tubulin-GFP-transfected KB cells. In **Figure 5a/e** the fluorescently-labeled microtubule network (green) of untreated KB cells is depicted for comparison. Already after 2 h of particle incubation (MSN-D3 labeled with Atto 633, red), endocytosis occurred to a high degree. Importantly, the microtubule network was still intact, suggesting that no release of Col had occurred at this point of time. A partial destruction of the microtubule network could be observed after 7 h (Figure 2c/g). After 22 h, cell death finally occurred which was indicated by vanishing of the tubulin structure, blurred green fluorescence signal and a spherical shape of the cells (for reference experiments see SI). Conclusively, the present cell experiments suggest a time-dependent intracellular release of immobilized MTS-Col from the mesoporous drug delivery vehicles. We anticipate that the acidification of the endosomal compartment while trafficking triggered the endosomal escape. The high buffering capacity of MSN-D3 caused high internal osmotic pressure subsequently leading to rupture of the endosomal membrane and access to the cytosol. Reducing agents present in the cytosol were able to cleave the disulfide bridges and Col has been efficiently released.

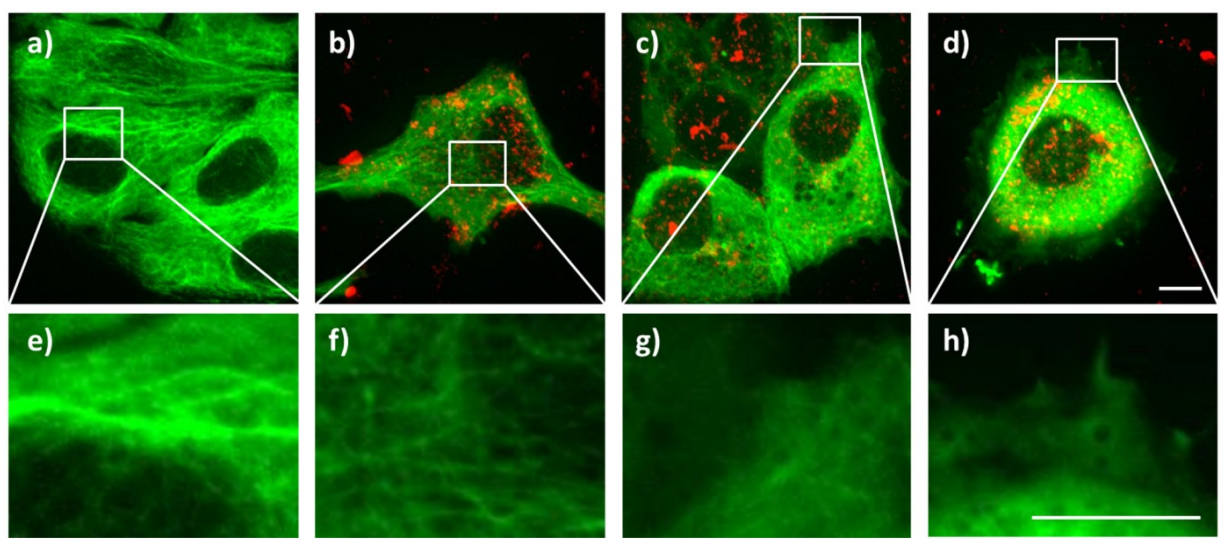

**Figure 5.** Fluorescence microscopy studies of a/e) untreated KB cells with GFP tagged tubulins (green). The cells were incubated with MSN-D3 loaded with immobilized colchicine (MTS-Col) inside the mesopores and labelled with Atto 633 (red) for b/f) 2 h, c/g) 7 h and d/h) 22 h. e-h) Zoom in on representative microtubule structures. A time-dependent destruction of the tubulin network can be observed, finally causing cell death. The scale bar represents 10 µm.

## 4. Conclusions

In conclusion, these *in vitro* release experiments with different cell lines and different cargo molecules incorporated into the mesoporous system of PAMAM dendron-coated silica nanoparticles suggest an intrinsic endosomal escape pathway followed by an intracellular redox-driven release of immobilized cargo molecules. Furthermore, they feature optional attachment of various cargos inside the mesopores via disulfide bridges and binding of different targeting ligands to the outer periphery of the nanoparticles which can be precisely tuned to target specific cancer cell lines. Cytotoxicity tests suggest a good bio-tolerability, since no adverse effects of the PAMAM dendron-coated MSNs on the metabolism of endothelial cells were observed. The combination of all these essential features into one multifunctional nanocarrier is anticipated to result in a powerful drug delivery system.

**Acknowledgements**

This work was funded by the Deutsche Forschungsgesellschaft (DFG) through the cluster Nanosystems Initiative Munich (NIM) and SFB 749, and the Center for NanoScience (CeNS). We thank Dr. Bastian Rühle for graphics design.

## Supporting Information

**Synthesis of propargyl-PAMAM dendrons**

Starting with propargylamine, the synthesis of PAMAM dendrons employs a Michael addition with methyl acrylate and amidation with 1,2-ethylendiamine in an alternating fashion. The methodology was developed by Lee *et al.* [1].

General procedure for Michael addition: A solution of the amino-terminal compound in anhydrous methanol (80 mL) was added drop-wise to a stirring solution of freshly distilled methyl acrylate in methanol over a period of approx. 1 h at 0 °C. The resulting solution was allowed to warm up to room temperature and stirred for 48 h. The reaction progress was monitored by NMR spectroscopy and stopped when the N-H signal had disappeared. Solvent and methyl acrylate were removed *in vacuo*, traces of methyl acrylate were removed by redissolving the residue in methanol and repeated removal of the solvent. Drying under high vacuum gave the desired pure product as orange viscous oil.

General procedure for amidation: A solution of the ester-terminal compound in methanol (80 mL) was added drop-wise to a stirring solution of freshly distilled 1,2-ethylendiamine in methanol over a period of approx. 1 h at 0 °C. The resulting solution was allowed to warm up to room temperature and stirred for 48 h. The reaction was monitored by NMR spectroscopy and stopped when no methyl ester signal could be detected anymore. Solvent was removed *in vacuo*, the excess of 1,2-ethylendiamine was removed successively by adding 100 mL of an azeotropic toluene/methanol mixture (9:1) and removal of the volatiles *in vacuo*. This procedure needed to be repeated four to six times with a final pressure after removal of < 10 mbar. Remaining toluene was removed by azeotropic distillation using methanol. Drying under high vacuum gave the desired pure product as orange viscous oil.

Dendron D0.5: Starting material: Propargylamine (1.50 g, 1.74 mL, 27.2 mmol, 1 eq.); Reagents: methyl acrylate (17.0 g, 17.9 mL, 197 mmol, 7.3 eq.) in 20 mL methanol; Yield: 5.74 g (25.3 mmol, 93 %); IR (film): 3280, 2956, 2840, 1741, 1641, 1542, 1437, 1203, 1175, 1056 cm$^{-1}$; $^1$H NMR (270 MHz, CDCl$_3$) d = 3.60 (s, 6H, OMe), 3.35 (d, $^4$J = 2.3 Hz, 2H, N-CH$_2$-C≡CH), 2.77 (t, J = 7.0 Hz, 4H, N-CH$_2$-CH$_2$), 2.39 (t, J = 7.1 Hz, 4H, N-CH$_2$-CH$_2$-CO), 2.14 (t, J = 2.3 Hz, 1H, CH$_2$-C≡C-H); $^{13}$C NMR (68 MHz, CDCl$_3$) d = 172.84 (C=O), 78.22 (CH$_2$-C≡CH), 73.47 (CH$_2$-C≡CH), 51.77 (NH$_2$-CH$_2$-CH$_2$), 49.13 (OMe), 42.10 (N-CH$_2$-C≡CH), 33.09 (N-CH$_2$-CH$_2$-CO) (NMR spectra are shown in Appendix 9.5); HRMS (ESI) (C$_{11}$H$_{17}$NO$_4$): m/z (%) = 228.1229 (100) [M+H]$^+$.

Dendron D1: Starting material: D0.5 (4.90 g, 21.6 mmol, 1 eq.); Reagents: 1,2-ethylendiamine (111 g, 100 mL, 1.85 mol, 86 eq.) in 30 mL methanol; Yield: 6.23 g (21.2 mmol, 98 %); IR (film): 3281, 3062, 2933, 2826, 1639, 1543, 1456, 1356, 1250, 1195, 935 cm$^{-1}$; $^1$H NMR (400 MHz, CDCl$_3$) d = 7.31 (s, 2H, NHCO), 3.37 (d, $^4J$ = 2.4 Hz, 2H, N-CH$_2$-C≡CH), 3.23 (dt, J = 6.0 Hz, 4.0 Hz, 4H, CONHCH$_2$-CH$_2$), 2.78 (t, J = 6.0 Hz, 4H, N-CH$_2$-CH$_2$), 2.76 (t, J = 5.6 Hz, 4H, N-CH$_2$-CH$_2$) 2.33 (t, J = 6.1 Hz, 4H, N-CH$_2$-CH$_2$-CONH), 2.19 (t, J = 2.4 Hz, 1H, CH$_2$-C≡C-H), 1.67 (br s, 4H, NH$_2$); $^{13}$C NMR (68 MHz, CDCl$_3$) d = 172.51 (C=O), 77.83 (CH$_2$-C≡CH), 73.85 (CH$_2$-C≡CH), 49.72 (N-CH$_2$-CH$_2$-CO), 42.22 (CONH-CH$_2$-CH$_2$), 41.80 (N-CH$_2$-C≡CH), 41.61 (NH$_2$-CH$_2$-CH$_2$), 34.23 (N-CH$_2$-CH$_2$-CO) (NMR spectra are shown in Appendix 9.5); HRMS (ESI) (C$_{13}$H$_{25}$N$_5$O$_2$): m/z (%) = 284.2081 (100) [M+H]$^+$, 306.1900 (93) [M+Na]$^+$, 142.6077 (86) [M+2H]$^{2+}$.

Dendron D1.5: Starting material: D1 (5.78 g, 20.4 mmol, 1 eq.); Reagents: methyl acrylate (28.5 g, 30.0 mL, 331 mmol, 16 eq.) in 40 mL methanol; Yield: 12.3 g (19.6 mmol, 96 %); IR (film): 3284, 2951, 2841, 1732, 1653, 1540, 1436, 1361, 1260, 1177, 841 cm$^{-1}$; $^1$H NMR (400 MHz, CDCl$_3$) d = 7.07 (s, 2H, NHCO), 3.63 (s, 12H, OMe), 3.42 (d, $^4J$ = 2.3 Hz, 2H, N-CH$_2$-C≡CH), 3.25 (dt, J = 6.0 Hz, 4.0 Hz, 4H, CONH-CH$_2$-CH$_2$), 2.81 (t, J = 6.6 Hz, 4H, N-CH$_2$-CH$_2$), 2.72 (t, J = 6.7 Hz, 8H, N-CH$_2$-CH$_2$-COOMe), 2.50 (t, J = 5.0 Hz, 4H, N-CH$_2$-CH$_2$), 2.39 (t, J = 6.6 Hz, 8H, N-CH$_2$-CH$_2$-COOMe), 2.34 (t, J = 6.5 Hz, 4H, N-CH$_2$-CH$_2$-CONH), 2.16 (t, J = 2.2 Hz, 1H, CH$_2$-C≡C-H); $^{13}$C NMR (68 MHz, CDCl$_3$) d = 172.91 (C=O), 171.81 (C=O), 77.86 (CH$_2$-C≡CH), 73.42 (CH$_2$-C≡CH), 52.82, 51.49, 49.27, 49.16, 41.00, 36.97, 33.70, 32.60 (NMR spectra are shown in Appendix 9.5); HRMS (ESI) (C$_{29}$H$_{49}$N$_5$O$_{10}$): m/z (%) = 628.3546 (100) [M+H]$^+$.

Dendron D2: Starting material: D1.5 (10.5 g, 16.7 mmol, 1 eq.); Reagents: 1,2-ethylendiamine (111 g, 100 mL, 1.85 mol, 110 eq.) in 20 mL methanol; Yield: 12.1 g (16.4 mmol, 98 %); IR (film): 3280, 2930, 2819, 1637, 1545, 1362, 1249, 1126, 933 cm$^{-1}$; $^1$H NMR (400 MHz, CDCl$_3$) d = 7.90 (s, 2H, NHCO), 7.68 (s, 4H, NHCO), 3.38 (d, $^4J$ = 2.3 Hz, 2H, N-CH$_2$, 3.25-3.15 (m, 12H), 2.78-2.73 (m, 12H), 2.67 (t, J = 6.0 Hz, 8H), 2.46 (t, J = 5.8 Hz, 4H), 2.34-2.25 (m, 12H), 2.19 (t, J = 2.0 Hz, 1H, CH$_2$-C≡C-H), 1.98 (br s, 8H, NH$_2$); $^{13}$C NMR (68 MHz, CDCl$_3$) d = 172.90 (C=O), 172.24 (C=O), 77.66 (CH$_2$-C≡CH), 73.66 (CH$_2$-C≡CH), 52.71, 49.67, 49.17, 44.48, 41.93, 41.24, 37.65, 34.15, 33.63 (NMR spectra are shown in Appendix 9.5); HRMS (ESI) (C$_{33}$H$_{65}$N$_{13}$O$_6$): m/z (%) = 740.5249 (100) [M+H]$^+$.

Dendron D2.5: Starting material: D2 (6.7 g, 10.4 mmol, 1 eq.); Reagents: methyl acrylate (85.5 g, 90.0 mL, 0.99 mol, 96 eq.) in 30 mL methanol; Yield: 14.5 g (10.2 mmol, 98 %); IR

(film): 3295, 2953, 2822, 1731, 1646, 1539, 1436, 1358, 1259, 1174, 844 cm$^{-1}$; $^1$H NMR (270 MHz, CDCl$_3$) d = 7.64 (s, 2H, NHCO), 7.01 (s, 4H, NHCO), 3.55 (s, 24H, OMe), 3.35 (d, $^4$J = 2.0 Hz, 2H, N-CH$_2$-C≡CH), 3.22-3.12 (m, 12H), 2.74-2.60 (m, 28H), 2.50-2.40 (m, 12H), 2.36-2.20 (m, 28H), 2.12 (t, J = 2.0 Hz, 1H, CH$_2$-C≡C-H); $^{13}$C NMR (68 MHz, CDCl$_3$) d = 172.99 (C=O), 172.89 (C=O), 172.30 (C=O), 78.01 (CH$_2$-C≡CH), 73.51 (CH$_2$-C≡CH), 52.90, 52.49, 51.59, 49.87, 49.71, 49.22, 40.96, 37.37, 37.14, 33.83, 32.57 (NMR spectra are shown in Appendix 9.5); HRMS (ESI) (C$_{65}$H$_{113}$N$_{13}$O$_{22}$): m/z (%) = 1462.7789 (60) [M+Cl]$^-$, 1428.8189 (15) [M+H]$^+$, 1450.8014 (6) [M+Na]$^+$.

Dendron D3: Starting material: D2.5 (6.42 g, 4.50 mmol, 1 eq.); Reagents: 1,2-ethylendiamine (122 g, 110 mL, 2.0 mol, 452 eq.) in 10 mL methanol; Yield: 6.5 g (3.91 mmol, 87 %); IR (film): 3271, 3074, 2933, 2863, 1636, 1540, 1436, 1359, 1197, 1127, 935 cm$^{-1}$; $^1$H NMR (270 MHz, CDCl$_3$) d = 7.98 (s, 2H, NHCO), 7.77 (s, 4H, NHCO), 7.57 (s, 8H, NHCO), 3.39 (d, $^4$J = 2.0 Hz, 2H, N-CH$_2$-C≡CH), 3.29-3.11 (m, 32H), 2.79-2.72 (m, 24H), 2.70-2.64 (m, 12H), 2.63-2.56 (m, 12H), 2.51-2.42 (m, 8H), 2.38-2.21 (m, 24H), 2.12 (t, J = 2.0 Hz, 1H, CH$_2$-C≡C-H), 1.65 (br s, 16H, NH$_2$); $^{13}$C NMR (68 MHz, CDCl$_3$) d = 173.37 (C=O), 173.15 (C=O), 172.82 (C=O), 77.43 (CH$_2$-C≡CH), 72.69 (CH$_2$-C≡CH), 53.05, 52.66, 52.17, 50.75, 50.25, 46.46, 45.01, 42.49, 42.32, 41.66, 37.86, 34.51, 34.13 (NMR spectra are shown in Appendix 9.5); HRMS (ESI) (C$_{73}$H$_{145}$N$_{29}$O$_{14}$): m/z (%) = 827.0835 (24) [M+2H]$^{2+}$, 1653.1595 (9) [M+H]$^+$.

**Synthesis of PAMAM Silane**

Huisgen azide-alkyne 1,3-dipolar cycloaddition with amino-terminated propargyl-PAMAM dendrons and AzTMS was performed based on a similar reaction of AzTMS with an alkyne compound reported by Lim *et al.* [2].

General procedure: The amino-terminated propargyl-PAMAM dendron **D1-3** (1 eq.) was dissolved in a suspension of molecular sieves (3 Å) in anhydrous methanol (15 mL) and DIPEA (2 eq.). The resulting suspension was degassed afterwards. Under an atmosphere of nitrogen, copper(I) iodide (10 mol%), as catalyst, and AzTMS (0.25 eq.) were added. The reaction was stirred for 24 h at room temperature and subsequently filtered using a syringe filter. After exposure to air the solution turned purple-blue very quickly. The solvent was removed *in vacuo*, drying of the crude product at high vacuum yielded a blue gum which was used without further purification.

S1: Starting material: **D1** (134 mg, 0.471 mmol); Reagents: AzTMS (24.2 mg, 24.2 mL, 0.118 mmol), DIPEA (122 mg, 164 mL, 0.942 mmol), CuI (9.50 mg, 0.050 mmol); Crude product: 204 mg containing 47.6 mg (106 mmol) of **S1.**

S2: Starting material: **D2** (1.48 g, 2.00 mmol); Reagents: AzTMS (103 mg, 103 mL, 0.500 mmol), DIPEA (517 mg, 697 mL, 4.00 mmol), CuI (38.1 mg, 200 mmol); Crude product: 1.38 g containing 425 mg (450 mmol) of **S2.**

S3**:** Starting material: **D3** (3.31 g, 2.00 mmol); Reagents: AzTMS (103 mg, 103 mL, 0.500 mmol), DIPEA (517 mg, 697 mL, 4.00 mmol), CuI (38.1 mg, 0.200 mmol); Crude product: 3.5 g containing 836 mg (450 mmol) of **S3**.

**Synthesis of dendron-functionalized MSNs (MSN-Dn, n = 1-3)**

Colloidal mesoporous silica nanoparticles (MSN-Dn, n = 1-3) were prepared according to a synthesis procedure published by Cauda *et al.* [3]. The amount of functionalized silane was calculated to be 1% of total silica. A mixture of TEA (14.3 g, 95.6 mmol), TEOS (1.56 g, 7.48 mmol) and MPTMS (92.3 mg, 87.3 mL, 0.47 mmol) was heated for 20 min under static conditions at 90 °C in a polypropylene reactor. Afterwards, a solution of cetyltrimethylammonium chloride (CTAC, 2.41 mL, 1.83 mmol, 25 wt% in $H_2O$) in $H_2O$ (21.7 g, 1.21 mmol) was preheated to 60 °C and added quickly to the TEOS solution. The reaction mixture was stirred vigorously (700 rpm) for 20 min while cooling down to room temperature. Subsequently, TEOS (138.2 mg, 0.922 mmol) was added in four equal increments every three minutes. The reaction was stirred for further 30 min. After this time, a mixture of TEOS (19.2 mg, 92.2 µmol) and functionalized trialkoxysilane (RTMS: **S1**, **S2**, or **S3**) (92.2 µmol) was added. For the PAMAM silanes **S1-3** (cf. SI), the amount of crude product to be used for particle functionalization was calculated from the percental mass of the silane in the crude product. Furthermore, the mixture of TEOS and PAMAM silane was dissolved in a solution of 2 mL methanol and 1 mL water briefly before the addition. The reaction was stirred at room temperature overnight. The suspension was diluted 1:1 with absolute ethanol, the colloidal MSNs were separated by centrifugation (19,000 rpm, 43,146 rcf, 20 min) and redispersed in absolute ethanol. The template extraction was performed by heating the samples under reflux at 90 °C (oil bath) for 45 min in a solution of ammonium nitrate (2 wt% in ethanol) followed by 45 min under reflux at 90 °C in a solution of 10 mL conc. HCl (37 %) in 90 mL ethanol. The extracted MSNs were collected by centrifugation after each extraction step and finally washed with 100 mL absolute ethanol. The resulting MSNs were stored in an ethanol/water solution (2:1).

**Synthesis of MSNs without dendron functionalization (MSN-D0)**

Core(-SH)-shell(-N$_3$) functionalized MSNs were synthesized as previously reported [3]. In brief, a mixture of TEA (14.3 g, 95.6 mmol), TEOS (1.56 g, 7.48 mmol) and MPTMS (92.3 mg, 87.3 mL, 0.47 mmol) was heated for 20 min under static conditions at 90 °C in a polypropylene reactor. Afterwards, a solution of cetyltrimethylammonium chloride (CTAC, 2.41 mL, 1.83 mmol, 25 wt% in H$_2$O) in H$_2$O (21.7 g, 1.21 mmol) was preheated to 60 °C and added quickly to the TEOS solution. The reaction mixture was stirred vigorously (700 rpm) for 20 min while cooling down to room temperature. Subsequently, TEOS (138.2 mg, 0.922 mmol) was added in four equal increments every three minutes. After another 30 min of stirring at room temperature, TEOS (19.3 mg, 92.5 µmol) and azidopropyl trimethoxysilane (AzTMS, 92.5 µmol) were added to the reaction. The resulting mixture was then allowed to stir at room temperature overnight. After addition of ethanol (100 mL), the MSNs were collected by centrifugation (19,000 rpm, 43,146 rcf for 20 min) and re-dispersed in absolute ethanol. The template extraction was performed by heating the MSN suspension under reflux (90 °C oil bath temperature) for 45 min in an ethanolic solution (100 mL) containing ammonium nitrate (NH$_4$NO$_3$, 2 g), followed by 45 min under reflux in a solution of concentrated hydrochloric acid (HCl, 10 mL) and absolute ethanol (90 mL). The mesoporous silica nanoparticles were collected by centrifugation and washed with absolute ethanol after each extraction step.

**Synthesis of amino-functionalized MSNs (MSN-NH$_2$)**

Core(-SH)-shell(-NH$_2$) functionalized MSNs were synthesized as previously reported [3]. In brief, a mixture of TEA (14.3 g, 95.6 mmol), TEOS (1.56 g, 7.48 mmol) and MPTMS (92.3 mg, 87.3 mL, 0.47 mmol) was heated for 20 min under static conditions at 90 °C in a polypropylene reactor. Afterwards, a solution of cetyltrimethylammonium chloride (CTAC, 2.41 mL, 1.83 mmol, 25 wt% in H$_2$O) in H$_2$O (21.7 g, 1.21 mmol) was preheated to 60 °C and added quickly to the TEOS solution. The reaction mixture was stirred vigorously (700 rpm) for 20 min while cooling down to room temperature. Subsequently, TEOS (138.2 mg, 0.922 mmol) was added in four equal increments every three minutes. After another 30 min of stirring at room temperature, TEOS (19.3 mg, 92.5 µmol) and aminopropyl triethoxysilane (APTES, 20.5 mg, 92.5 µmol) were added to the reaction. The resulting mixture was then allowed to stir at room temperature overnight. After addition of ethanol (100 mL), the MSNs were collected by centrifugation (19,000 rpm, 43,146 rcf for 20 min) and re-dispersed in absolute ethanol. The template extraction was performed by heating the MSN suspension under reflux (90 °C oil bath temperature) for 45 min in an ethanolic solution (100 mL) containing ammonium nitrate ($NH_4NO_3$, 2 g), followed by 45 min under reflux in a solution of concentrated hydrochloric acid (HCl, 10 mL) and absolute ethanol (90 mL). The mesoporous silica nanoparticles were collected by centrifugation and washed with absolute ethanol after each extraction step.

**Further characterization results**

The synthesis of PAMAM dendron-coated MSNs was carried out in three steps. First, propargyl-PAMAM dendrons were obtained in three different generations (D1, D2, and D3) by iterative steps of Michael additions and amidations following a previously described procedure [2]. The synthetic route comprising alternating use of the reagents methyl acrylate and 1,2-ethylenediamine for a stepwise creation of increasing PAMAM dendron generations is shown in Figure S7. For a convergent synthesis route to covalently attach the PAMAM dendrons at the external MSN surface, a PAMAM dendron silane precursor was synthesized in a second step. Three PAMAM dendron silanes were synthesized by performing a Huisgen click reaction to attach azidopropyl trimethoxysilane to the propargyl-PAMAM dendrons. According to a previous report [4], the azidopropyl trimethoxysilane was synthesized and subsequently reacted with alkyne-PAMAM dendrons to gain triazole-linked PAMAM moieties by using azide-alkyne click chemistry. Successful conversion was observed by IR spectroscopy (Figure S14). In a last step, this silane linker was used in a delayed co-condensation approach to create core-shell functionalized MSNs via a sol-gel process, as described previously [3]. In brief, tetraethyl orthosilicate (TEOS) was used as a silica source and cetyltrimethylammonium chloride (CTAC) as template. The silica framework formation was catalyzed by the base triethanolamine (TEA), which also served as a structure directing agent to result in colloidal silica nanoparticles. Specifically, three different bifunctional MSNs (MSN-Dn, n = 1-3) equipped with different dendron generations (D1, D2 and D3) on the external surface were prepared via a delayed co-condensation approach. The MSNs additionally consisted of a thiol-functionalized particle core, and additionally PAMAM dendrons were exclusively present on the outer particle surface.

**Nitrogen sorption measurements.** For all resulting MSN samples, the nitrogen sorption measurements showed type IV isotherms indicating a mesoporous structure of the nanoparticles with typical inflection points at about 0.3 $p/p_0$ (Figure S1). At higher partial pressures (about 0.9 $p/p_0$) a second hysteresis loop arose for all samples, which we attribute to interparticle textural porosity. For the reference sample MSN-D0 a relatively high BET surface area and pore volume was observed (data summarized in Table S1). For all PAMAM dendron-coated MSNs these porosity parameters decreased depending on the size of the dendrons. This effect was partially due to the increasing sample mass by addition of the non-porous organic polymer. Furthermore, large PAMAM dendron generations might have caused clogging of some pore entrances towards to access of nitrogen molecules at the low measurement temperatures (-196 °C). A slight average pore constriction could also be observed, which correlates with the size of the attached PAMAM structures. This decrease could be due to the reduction of the pore mouth diameters partially covered by frozen organic PAMAM moieties. Nevertheless, PAMAM dendron-coated MSNs still exhibited a large accessible mesoporous structure offering enough space for the incorporation of cargo molecules.

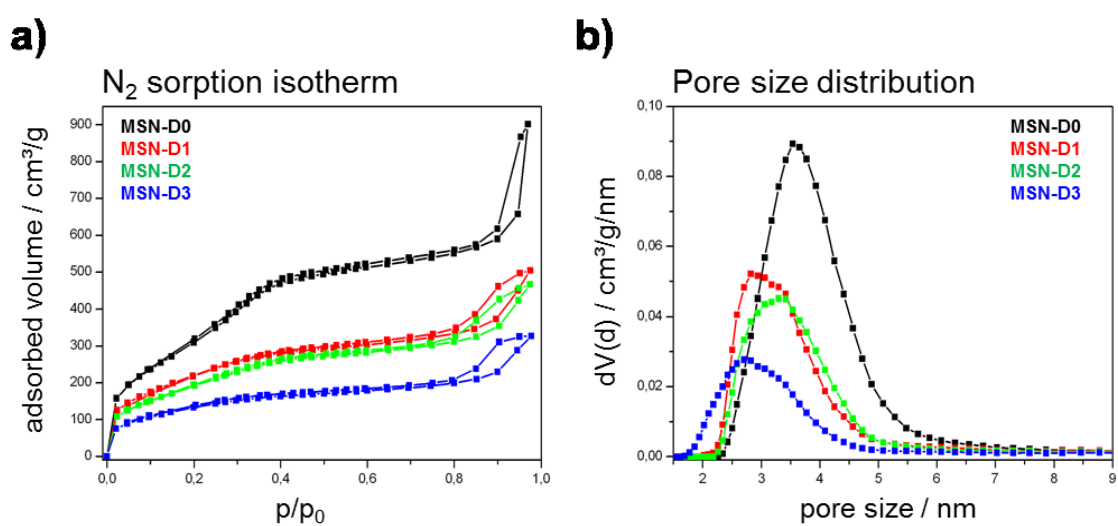

**Figure S1.** a) Nitrogen sorption isotherms and b) DFT pore size distributions of functionalized MSNs. MSN-D0 (black), MSN-D1 (red), MSN-D2 (green), and MSN-D3 (blue).

**Table S1.** Structural parameters of functionalized MSNs.

| Sample | BET surface area [m²/g] | Pore volume[a] [cm³/g] | DFT pore size[b] [nm] |
|---|---|---|---|
| MSN-D0 | 1190 | 0.74 | 2.9 – 4.4 |
| MSN-D1 | 823 | 0.42 | 2.5 – 3.9 |
| MSN-D2 | 718 | 0.39 | 2.6 – 4.1 |
| MSN-D3 | 497 | 0.24 | 2.2 – 3.8 |

a) Pore volume is calculated up to a pore size of 8 nm to remove the contribution of the interparticle porosity; b) DFT pore size refers to FWHM of the corresponding pore size distribution.

**Transmission electron microscopy.** Transmission electron microscopy shows spherically shaped nanoparticles with sizes of about 70 nm in diameter for PAMAM dendron-coated MSNs (Figure S2c, MSN-D3). A worm-like porous structure consisting of radially grown mesoporous channels was present for all samples. The external organic shell consisting of PAMAM dendrons cannot be visualized with conventional TEM due to lack of sufficient contrast against the carbon-coated sample holder (copper grid).

**Dynamic light scattering.** Additionally, dynamic light scattering (DLS) measurements were performed to determine particle sizes for all samples; the particle size distributions are shown in Figure S2a. All samples featured narrow size distributions with an average particle size of about 122 nm. The size difference between TEM and DLS data is attributed to a tendency for weak agglomeration in solution. Summarizing, the above data confirm the formation of well-defined mesoporous nanoparticles with good colloidal stability in aqueous solution.

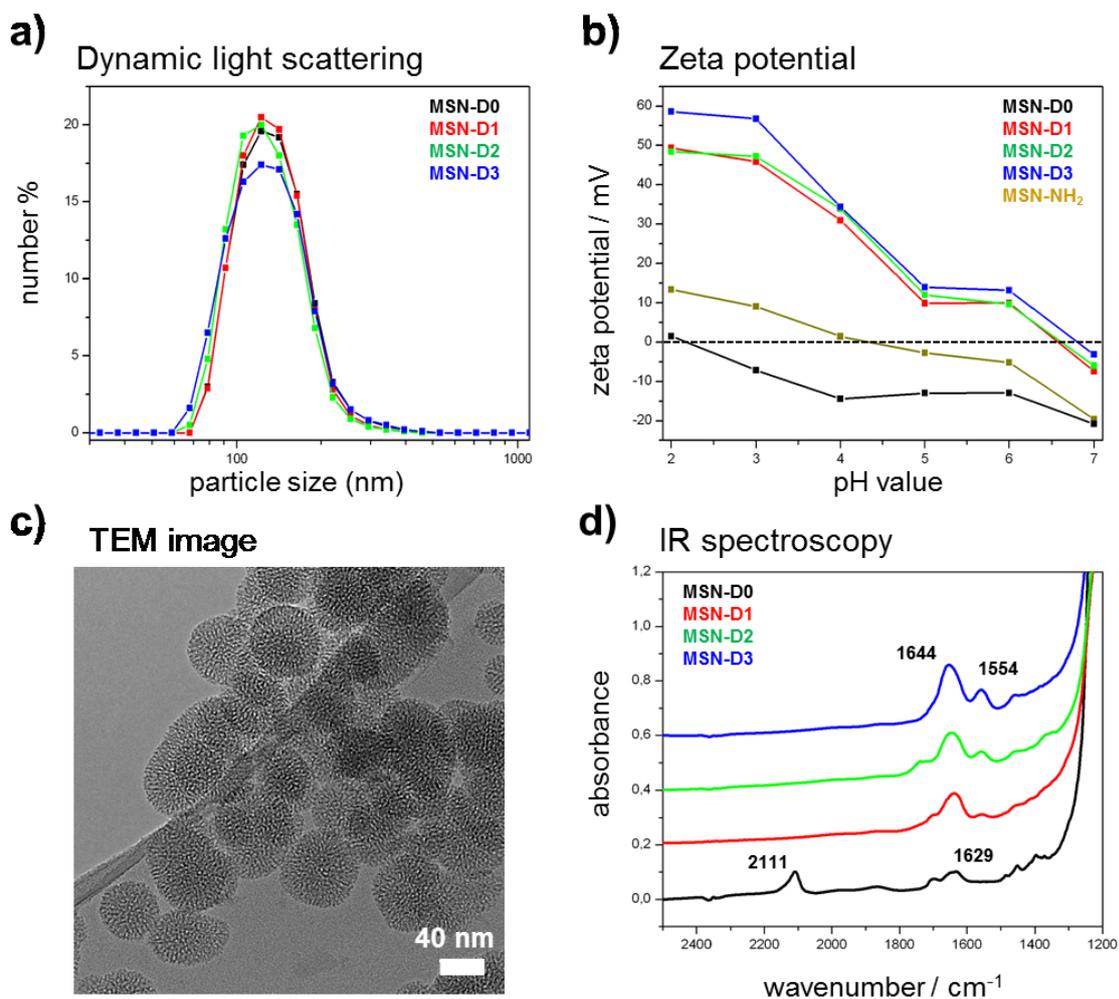

**Figure S2.** a) Dynamic light scattering (DLS) measurements, b) zeta potential measurements, c) transmission electron micrograph, and d) IR spectroscopy of functionalized MSNs. MSN-D0 (black), MSN-D1 (red), MSN-D2 (green), MSN-D3 (blue), and MSN-NH$_2$ (brown). The TEM image shows MSN-D3. For clarity reasons, the IR spectra are shifted along the y-axis by 0.2 units.

**Zeta potential measurements.** Zeta potential measurements showed drastic changes in the surface charge of PAMAM dendron-coated MSNs (Figure S2b). At acidic pH values, highly positive surface charges were observed for all three samples containing PAMAM dendrons at the outer particle surface. A zeta potential of about +60 mV at pH 2 was obtained for the sample MSN-D3 consisting of the largest PAMAM generation. This sample exhibited the highest amino group content leading to a highly positively charged particle surface. The samples MSN-D1 and MSN-D2 showed slightly smaller zeta potential values (about +50 mV at pH 2). Both samples showed an almost identical zeta potential curve in acidic milieu. This suggests that the content of accessible amino groups on the periphery of both particle types was nearly identical. It is reasonable to expect a higher density of $1^{st}$ generation PAMAM dendrons on the external particle surface compared to the more bulky $2^{nd}$ generation due to steric hindrance of the larger molecules. Nevertheless, the PAMAM dendron-coated MSNs showed a significant difference of the zeta potential values compared to the reference sample MSN-D0. Here, the isoelectric point (IEP) was close to pH 2 which resulted in a negatively charged particle surface over the pH range measured. Furthermore, MSNs functionalized with aminopropyl groups (MSN-NH$_2$) featured only slightly positive surface charge with an IEP of about 4.2. The IEPs for the PAMAM-derived particles were significantly shifted to higher pH values (about 6.7). These highly positively charged particle surfaces in acidic milieu gave evidence for a high proton acceptor density of the polymer shells resulting in a high buffering capacity.

**Titration experiments.** For further investigation of the buffering capacity of PAMAM dendron-coated MSNs, all samples were titrated against an aqueous solution of NaOH (0.01 M). As depicted in Figure S3, PAMAM-derived samples featured a significant increase in the required volume of NaOH solution to be neutralized (pH 7). This suggested a high tendency for proton uptake, and a stepwise increase in buffering capacity was observed for higher PAMAM dendron generations. MSN-D3 provided great potential to act like a proton sponge showing optimal buffering effect at around pH 6 which perfectly fits the endosomal acidification range. In general, a high buffering capacity in the physiological range (pH 5 – 7) is desirable to enable endosomal release via the proton sponge mechanism. For PAMAM dendrimer or dendron structures, the buffering capacity is mainly provided by the tertiary amino groups [5, 6].

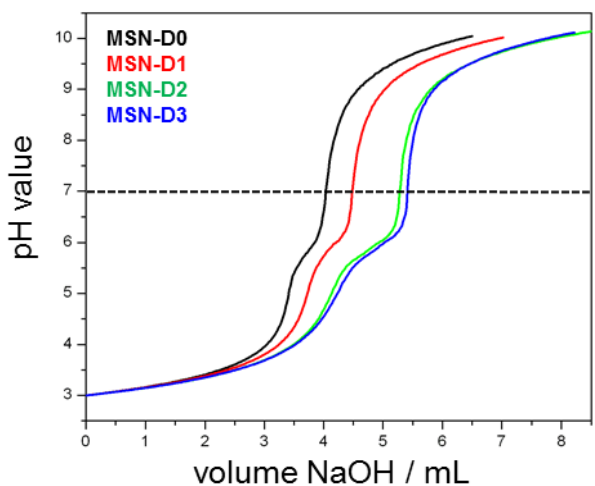

**Figure S3.** Titration data of functionalized MSNs. MSN-D0 (black), MSN-D1 (red), MSN-D2 (green), and MSN-D3 (blue).

**IR spectroscopy.** Organic functional groups present on the MSNs were examined with IR spectroscopy (Figure S2d, for full range IR spectra see also Figure S15). All samples featured typical bands of the silica framework appearing at 1240 – 1050 cm$^{-1}$ (asymmetric stretching vibration of Si-O-Si), and at 964 and 796 cm$^{-1}$ (asymmetric bending and stretching vibration of Si-OH). The reference sample MSN-D0 showed a characteristic signal at 2111 cm$^{-1}$, which was related to the azide vibration mode (asymmetric stretching vibration of –N=N+=N-). Additionally, a band at 1629 cm$^{-1}$ indicated the bending modes of physisorbed water. This band was present in all spectra but was partially covered by other, more intensive bands for the PAMAM dendron-coated particles. The nanoparticles MSN-D1, MSN-D2 and MSN-D3 showed two additional bands at 1644 cm$^{-1}$ (C=O stretching vibration) and 1554 cm$^{-1}$ (N-H deformation and C-N stretching vibration) which were attributed to the amide bonds of the PAMAM dendron moieties. An increase in intensity of these signals was observed with increasing PAMAM dendron generations indicating increasing amounts of functional groups within the samples.

**CHN analysis.** CHN elemental analysis provided quantitative information about the degree of organic functionalization of the modified MSNs. The data are shown in Table S2 (aminopropyl functionalized MSNs (MSN-NH$_2$) have been used as a reference). The nitrogen content (wt% N) of the samples correlates with the amount of PAMAM functionalization A stepwise increase in nitrogen content for samples with increasing PAMAM generations was observed. Furthermore, the measured values for all PAMAM dendron coated MSNs were significantly higher compared to the reference samples MSN-D0 and MSN-NH$_2$. These high nitrogen contents are the basis for a high protonation level in acidic milieu.

**Table S2.** CHN elemental analysis data for functionalized MSNs.

| Sample | MSN-D0 | MSN-D1 | MSN-D2 | MSN-D3 | MSN-NH$_2$ |
|---|---|---|---|---|---|
| Wt% N | 0.860 | 1.42 | 2.09 | 2.47 | 0.350 |

**Solid state NMR measurements.** Further characterization of the attached functional groups was performed by $^{13}$C solid state NMR spectroscopy (Figure S4). MSN-D0 showed distinct signals for the propyl chains of the mercaptopropyl- and azidopropyl-moieties (which have been incorporated into the silica framework during the co-condensation approach) at 53 ppm (a, -$\underline{C}$H$_2$-N$_3$), 27 ppm (b, -$\underline{C}$H$_2$-SH), 22 ppm (c, CH$_2$-$\underline{C}$H$_2$-CH$_2$-), and 10 ppm (d, Si-$\underline{C}$H$_2$-). Additional peaks of high intensity at 58 ppm (e, O-$\underline{C}$H$_2$-CH$_3$) and 15 ppm (f, O-CH$_2$-$\underline{C}$H$_3$) were present in all samples and were attributed to surface-bound ethoxy groups resulting from the extraction steps in ethanolic solution. PAMAM dendron functionalized MSNs featured characteristic peaks for the amide groups of the PAMAM dendrons at 173 ppm (g, C=O). Furthermore, weak signals at 144 ppm (h) and 125 ppm (i) were observed, these are attributed to the two carbon atoms in the triazole ring, thus providing evidence for a successful click reaction of the PAMAM dendron silane precursor. Several strong signals in the range between 60 to 10 ppm correspond to different types of methylene groups belonging to the PAMAM moieties (52 and 40 ppm (k and l, N-$\underline{C}$H$_2$-R), 21 and 10 ppm (m and n, R-$\underline{C}$H$_2$-R)).

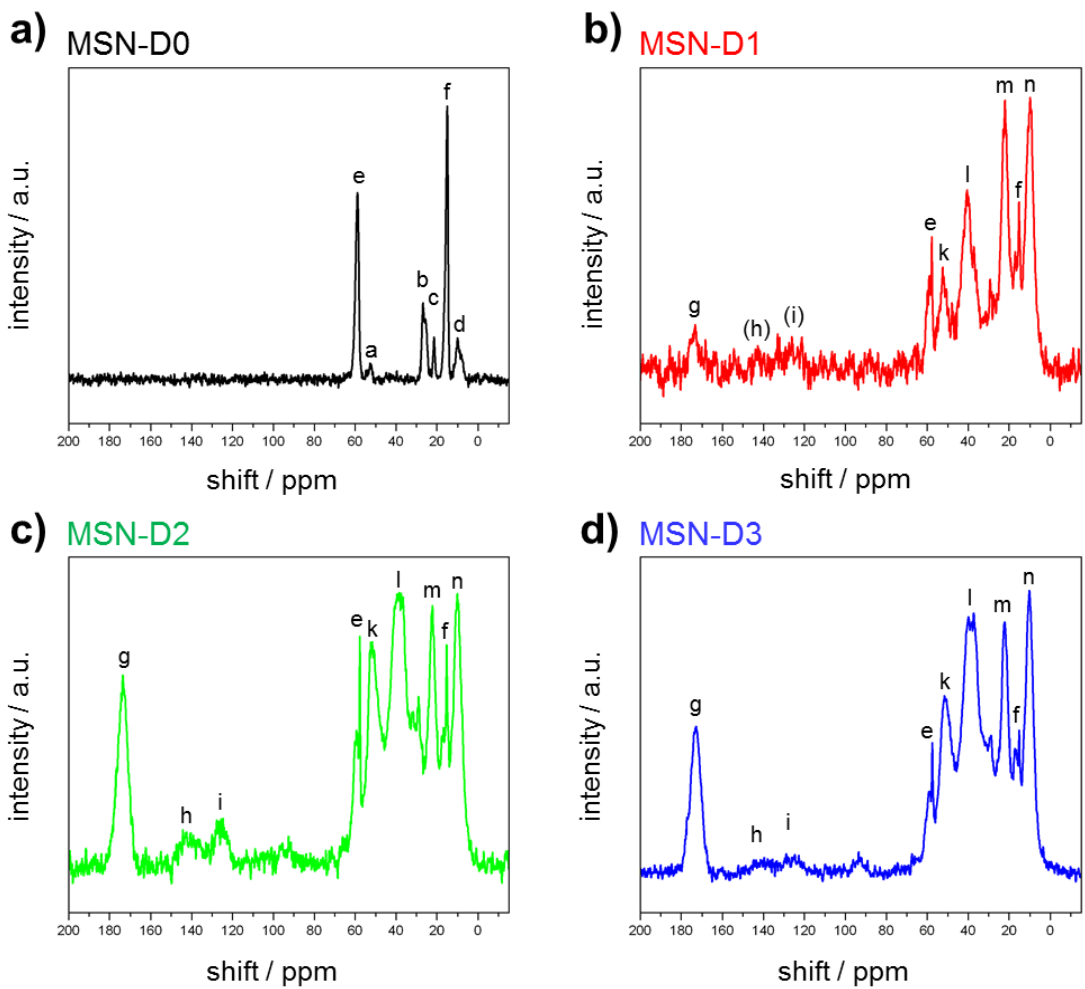

**Figure S4.** $^{13}$C solid state NMR spectra of functionalized MSNs. a) MSN-D0 (black), b) MSN-D1 (red), c) MSN-D2 (green), and d) MSN-D3 (blue). a refers to -$\underline{C}H_2$-$N_3$ (53 ppm), b to -$\underline{C}H_2$-SH (27 ppm), c to $CH_2$-$\underline{C}H_2$-$CH_2$- (22 ppm), d to Si-$\underline{C}H_2$- (10 ppm), e to O-$\underline{C}H_2$-$CH_3$ (58 ppm), f to O-$CH_2$-$\underline{C}H_3$ (15 ppm), g to C=O (173 ppm), h and i to carbon atoms of the triazole ring (144 and 125 ppm, respectively), k and l to N-$\underline{C}H_2$-R (52 and 40 ppm, respectively), and m and n to R-$\underline{C}H_2$-R (21 and 10 ppm, respectively).

From all the above results, we conclude a successful synthesis of core-shell functionalized MSNs with different generations of PAMAM dendrons (D1, D2 and D3) via the delayed co-condensation approach. The PAMAM moieties are exclusively located at the external particle surface resulting in an organic polymer coating of MSNs featuring high buffering capacity. The sample MSN-D3 showed optimal properties and is expected to have great potential for generating the proton sponge effect. Hence, in the main text we exclusively investigated the MSN-D3 particles for cargo release, cell uptake and cell targeting experiments.

**Further cell experiments, Images**

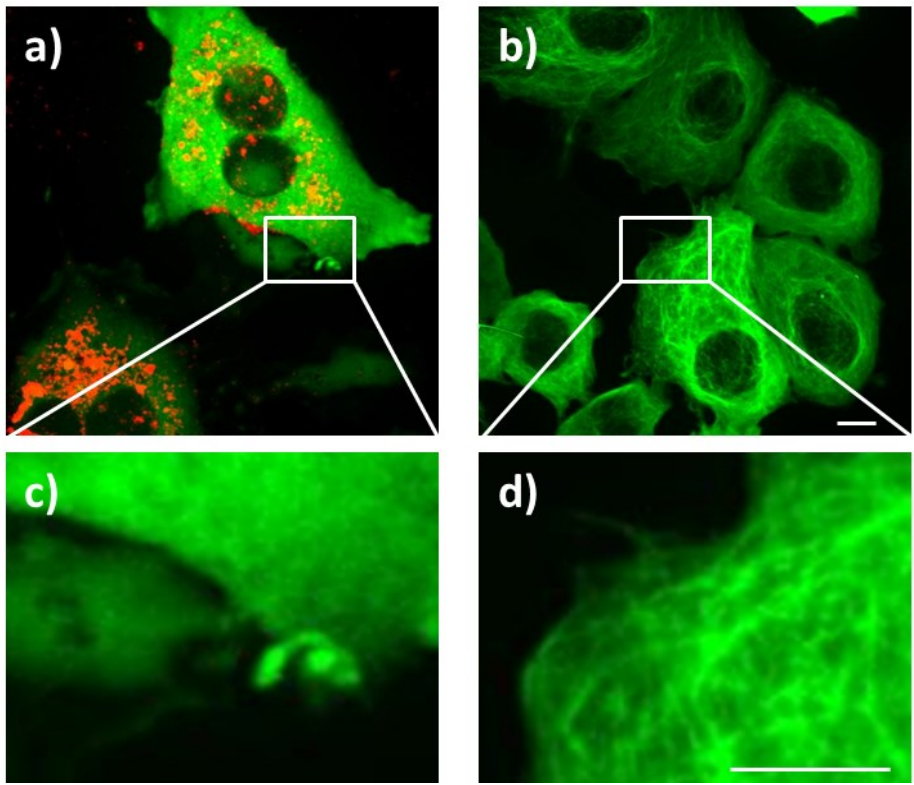

**Figure S5.** Fluorescence microscopy of KB cells with GFP tagged tubulins (green). a/c) The cells were incubated with MSN-D3 loaded with covalently attached colchicine (MTS-Col) in the mesopores and labeled with Atto 633 (red) after 22 h incubation on the cells. b/d) Cells are incubated with sample supernatant after particle separation by centrifugation for 22h. c-d) Zoom in on representative microtubule structures. A destruction of the tubulin network can only be observed for particle-incubated cells. The scale bar represents 10 µm.

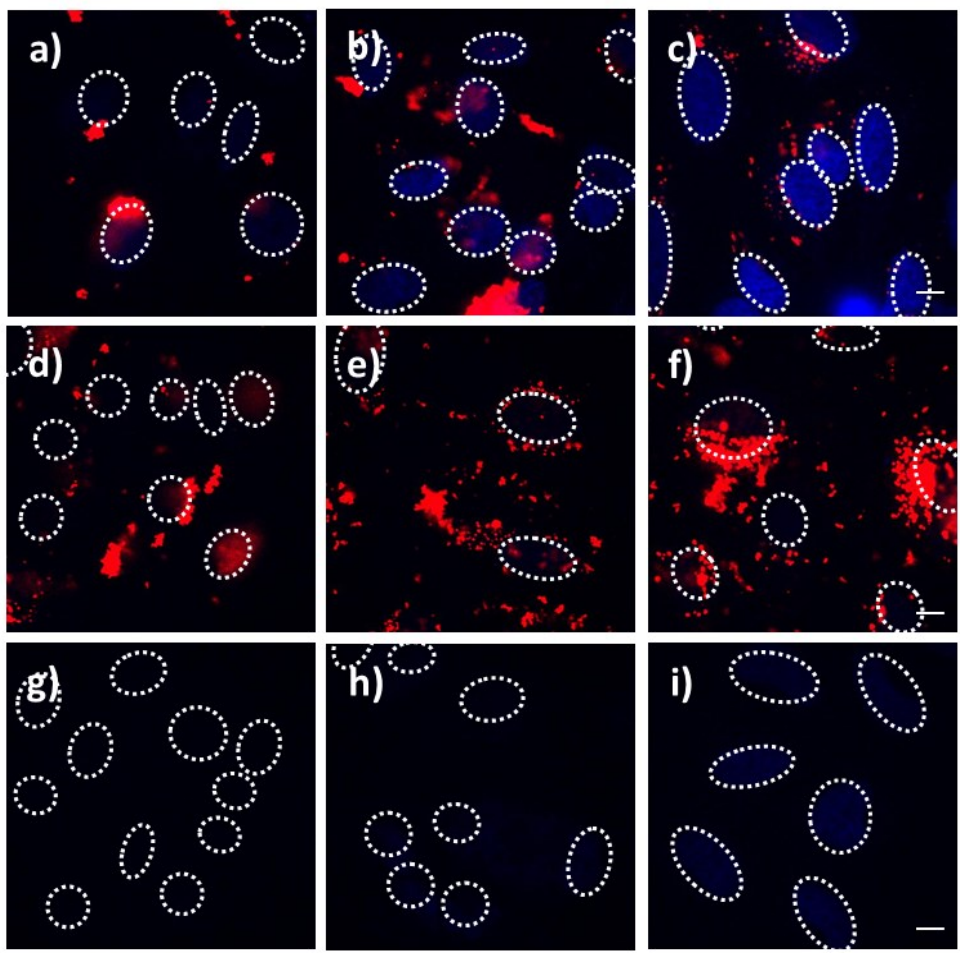

**Figure S6.** Fluorescence microscopy of HeLa cells incubated with MSN-D3-MTS-DAPI (a, b, c), MSN-NH$_2$-MTS-DAPI (d, e, f), and the supernatant of MSN-D3-MTS-DAPI (g, h, i). The nanoparticles have been loaded with covalently attached DAPI (MTS-DAPI, blue) and labeled with Atto 633 (red). The cells have been incubated with the samples for 5 h (a, d, g), 24 h (b, e, h) and 61 h (c, f, i). The nuclei are indicated with dashed circles. A time-dependent nuclei staining can be observed for MSN-D3-MTS-DAPI, suggesting triggered release of the cargo molecules from the nanocarriers after endosomal escape conceivably via the proton sponge effect and excess to the cytosol. For MSN-NH$_2$-MTS-DAPI and the supernatant incubation almost no nucleic staining can be detected which implies that no DAPI has been release form the particles or has been present in the solution, respectively. The scale bars represent 10 μm.

**Synthesis Scheme for PAMAM Dendrons**

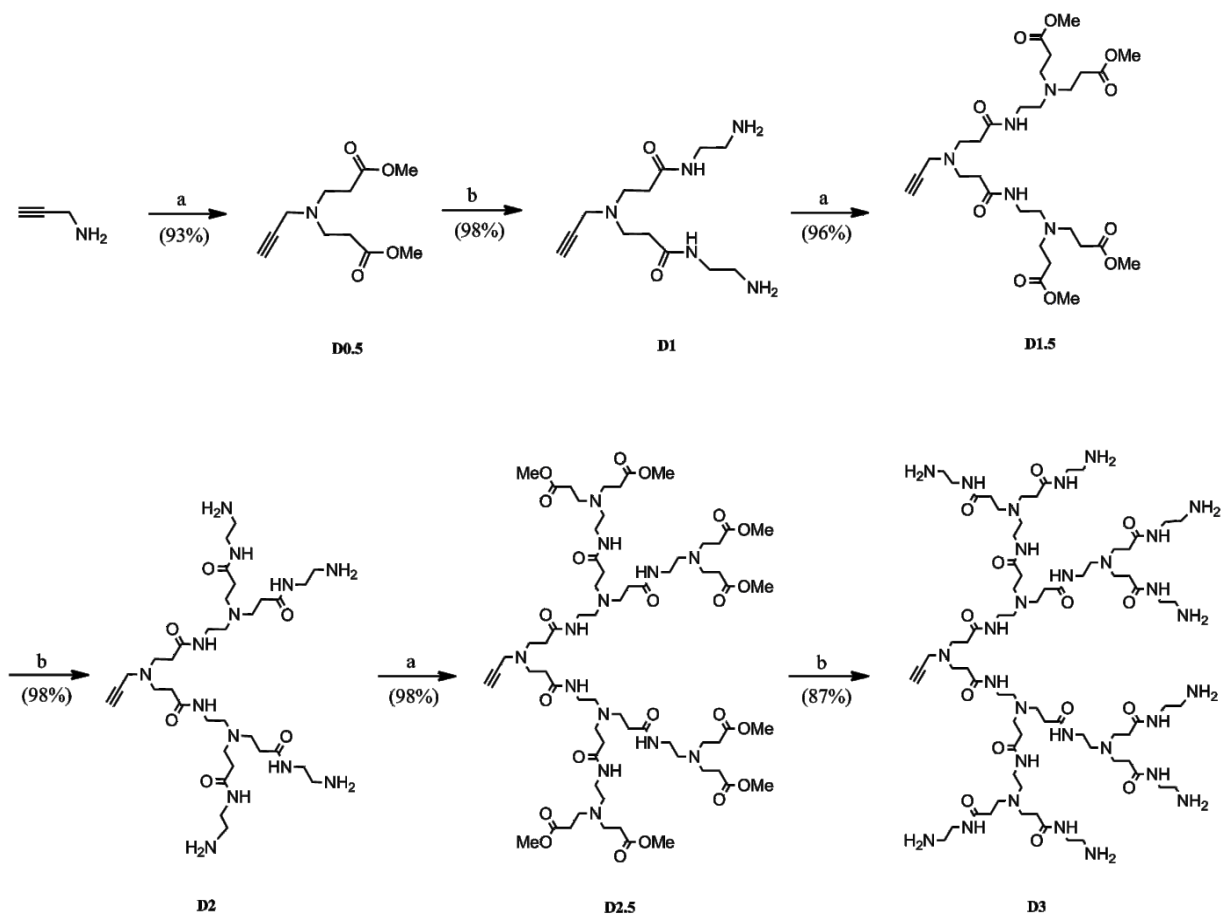

**Figure S7.** Synthesis of propargyl-PAMAM dendrons (D0.5 – D3). Reagents and reaction conditions: a) methyl acrylate (large excess), MeOH, 0 → 25 °C, 48 h and b) 1,2-ethylenediamine (large excess), MeOH, 0 → 25 °C, 48 h.

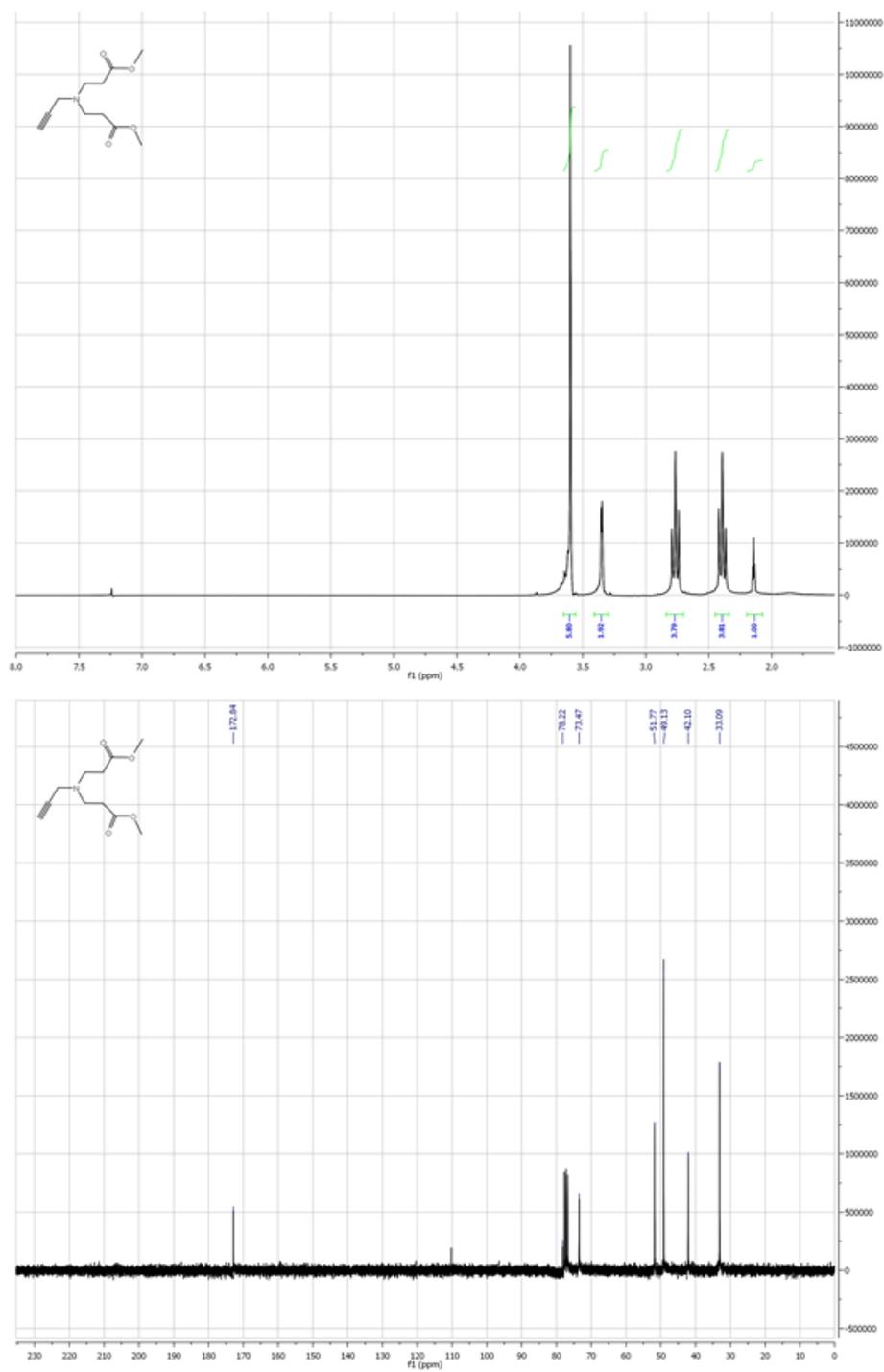

**Figure S8.** $^1$H and $^{13}$C NMR spectrum of propargyl-dendron D0.5.

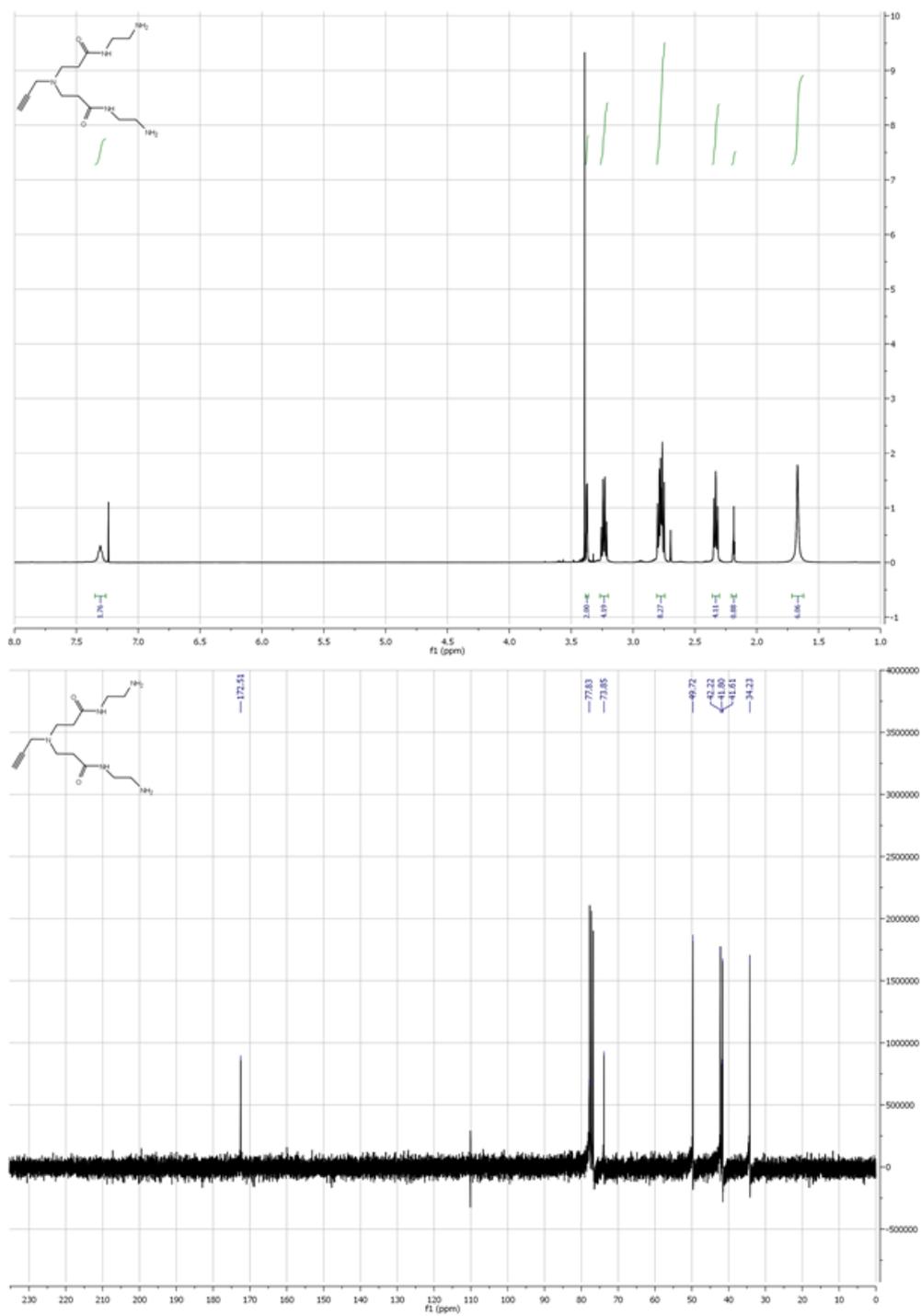

**Figure S9.** $^1$H and $^{13}$C NMR spectrum of propargyl-dendron D1.

**Figure S10.** $^1$H and $^{13}$C NMR spectrum of propargyl-dendron D1.5.

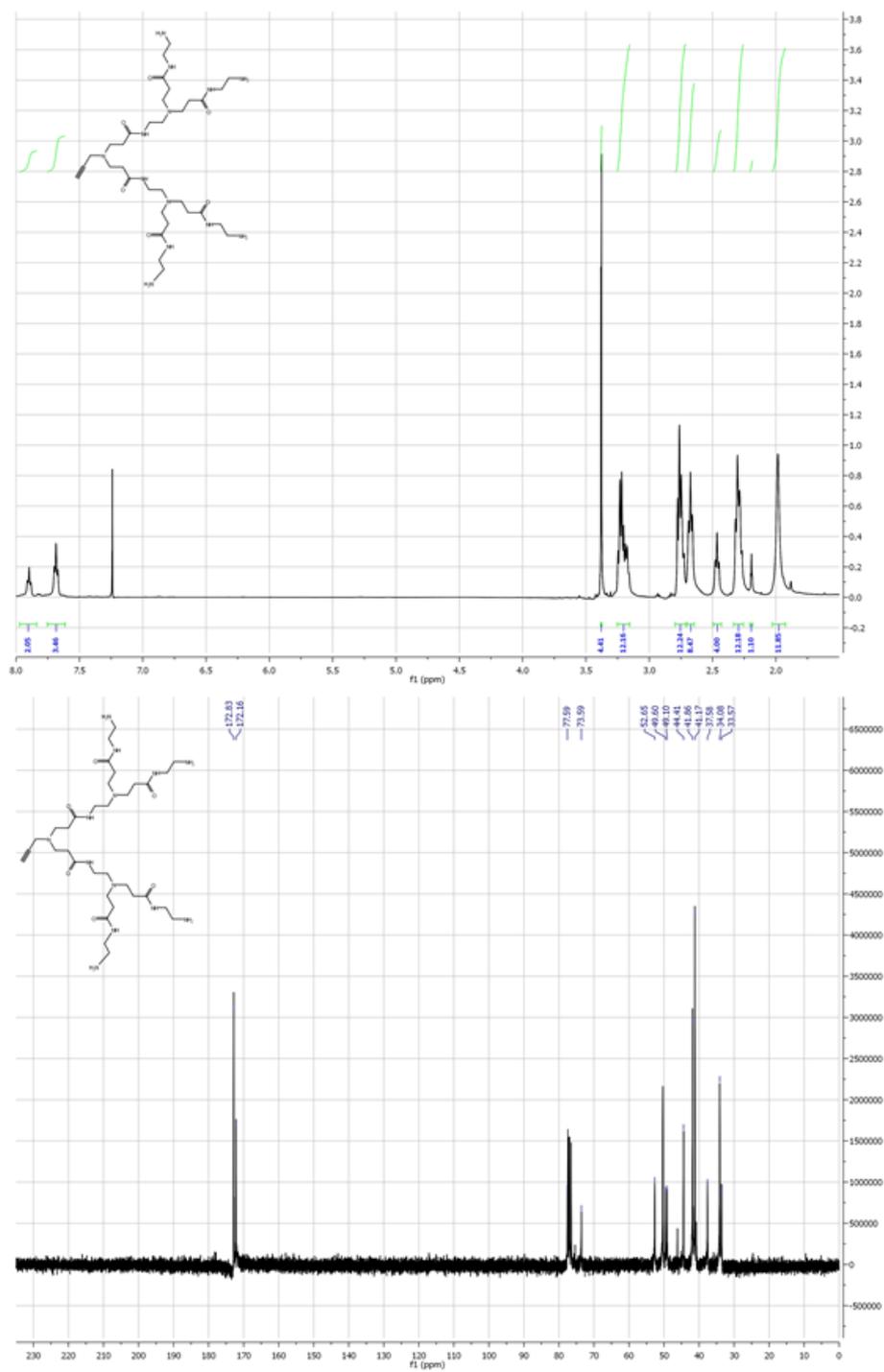

**Figure S11.** $^1$H and $^{13}$C NMR spectrum of propargyl-dendron D2.

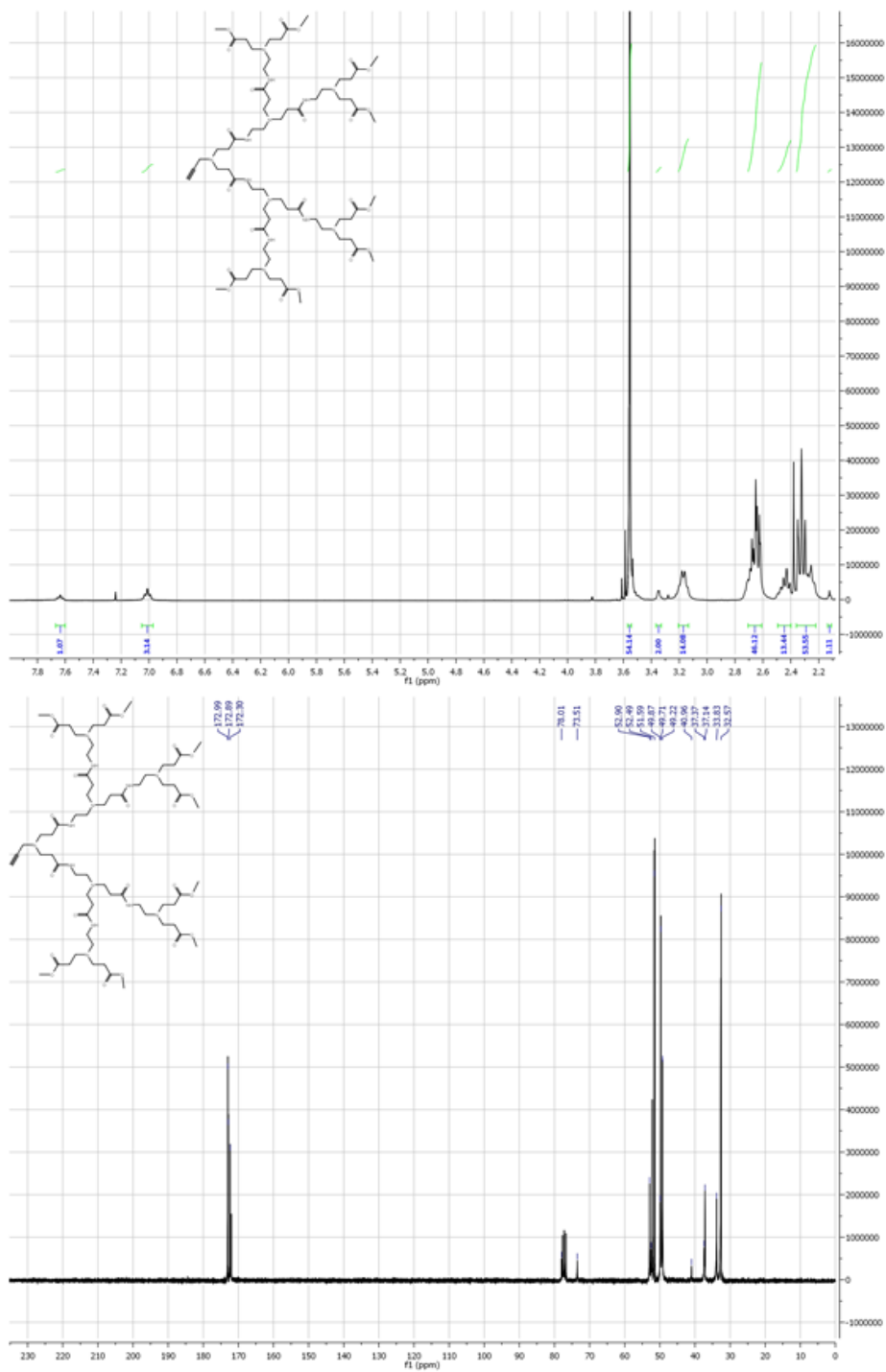

**Figure S12.** $^1$H and $^{13}$C NMR spectrum of propargyl-dendron D2.5.

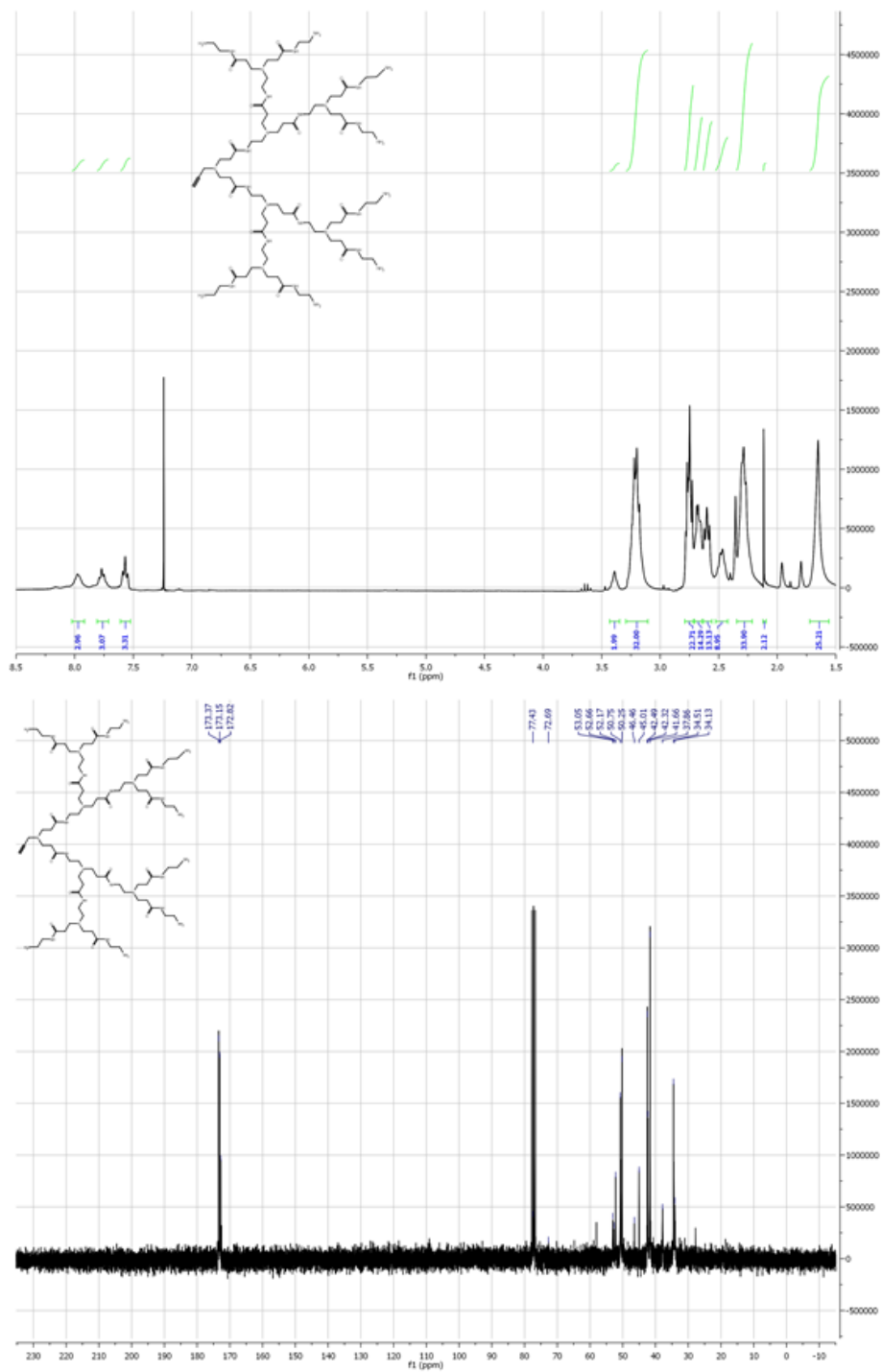

**Figure S13.** $^1$H and $^{13}$C NMR spectrum of propargyl-dendron D3.

**IR Spectra**

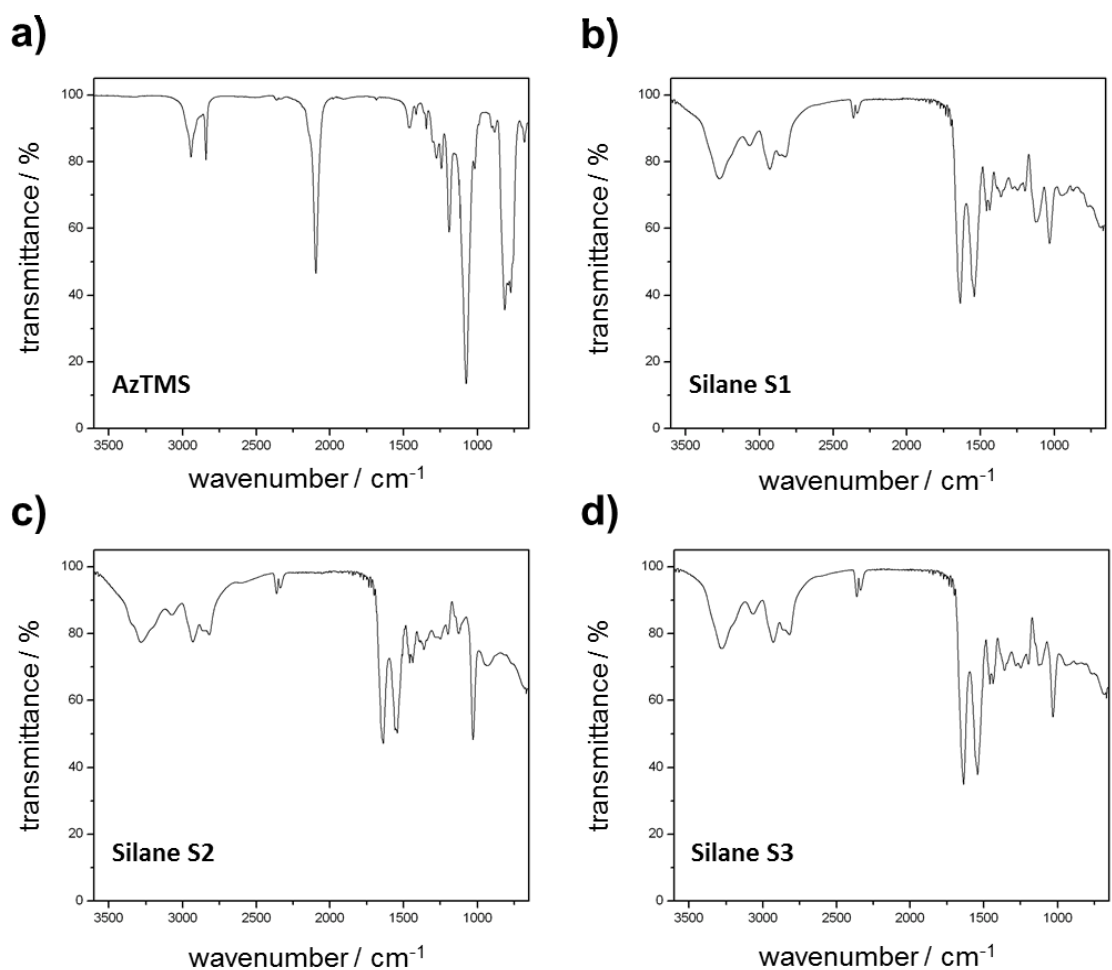

**Figure S14.** IR spectroscopy data of a) AzTMS, b) dendron-silane S1, c) dendron-silane S2, and d) dendron-silane S3.

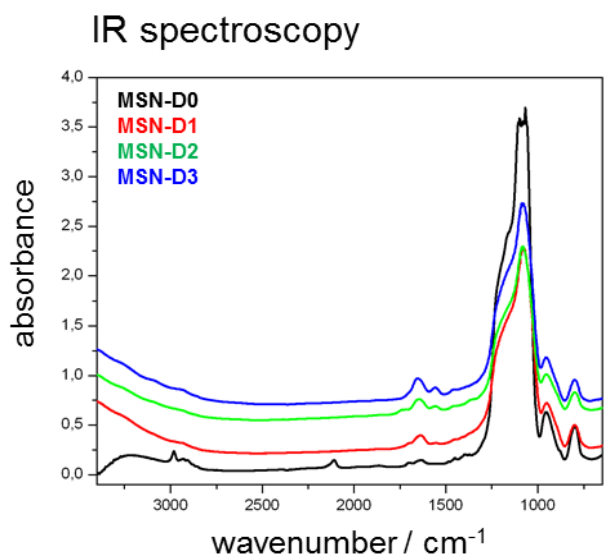

**Figure S15.** IR spectroscopy data of functionalized MSNs.

**References Supporting Information**